\documentclass[%
 aip,
 amsmath,amssymb,
preprint,%
]{revtex4-1}

\usepackage{graphicx}
\usepackage{dcolumn}
\usepackage{bm}
\usepackage[utf8]{inputenc}
\usepackage[T1]{fontenc}
\usepackage{mathptmx}
\usepackage{amsmath}
\usepackage{amssymb}
\usepackage{etoolbox}
\usepackage{yfonts}
\usepackage{xcolor}
\usepackage{ulem}
\usepackage{siunitx} 
\usepackage{physunits} 
\usepackage[colorlinks=true, allcolors=blue]{hyperref}
\usepackage{hyperref} 

\newcommand{\commentout}[1]{}

\usepackage{soul} 

\newcommand*\chem[1]{\ensuremath{\mathrm{#1}}} 

\newcommand{\kb}{\ensuremath{k_\mathrm{B}}}
\newcommand{\Ns}{\ensuremath{N_\mathrm{spec}}}
\newcommand{\Nr}{\ensuremath{N_\mathrm{rxn}}}

\DeclareMathAlphabet\mathbfcal{OMS}{cmsy}{b}{n}

\makeatletter
\def\@email#1#2{%
 \endgroup
 \patchcmd{\titleblock@produce}
  {\frontmatter@RRAPformat}
  {\frontmatter@RRAPformat{\produce@RRAP{*#1\href{mailto:#2}{#2}}}\frontmatter@RRAPformat}
  {}{}
}%
\makeatother
\begin{document}

\preprint{AIP/123-QED}

\title{Thermodynamic consistency and fluctuations in mesoscopic stochastic simulations of reactive gas mixtures}

\author{Matteo Polimeno}
\affiliation{Department of Applied Mathematics, University of California, Merced, California 95343, USA}
\author{Changho Kim}
\email{ckim103@ucmerced.edu}
\affiliation{Department of Applied Mathematics, University of California, Merced, California 95343, USA}
\author{Fran\c{c}ois Blanchette}
\affiliation{Department of Applied Mathematics, University of California, Merced, California 95343, USA}
\author{Ishan Srivastava}
\affiliation{Center for Computational Sciences and Engineering, Lawrence Berkeley National Laboratory, Berkeley, California 94720, USA}
\author{Alejandro L. Garcia}
\affiliation{Department of Physics and Astronomy, San Jose State University, San Jose, California 95192, USA}
\author{Andy J. Nonaka}
\affiliation{Center for Computational Sciences and Engineering, Lawrence Berkeley National Laboratory, Berkeley, California 94720, USA}
\author{John B. Bell}
\affiliation{Center for Computational Sciences and Engineering, Lawrence Berkeley National Laboratory, Berkeley, California 94720, USA}

\date{\today}

\begin{abstract}
It is essential that mesoscopic simulations of reactive systems reproduce the correct statistical distributions at thermodynamic equilibrium.
By considering a compressible fluctuating hydrodynamics (FHD) simulation method of ideal gas mixtures undergoing reversible reactions described by the chemical Langevin equations, we show that thermodynamic consistency in reaction rates and the use of instantaneous temperatures for the evaluation of reaction rates is required for fluctuations for the overall system to be correct.
We then formulate the required properties of a thermodynamically-consistent reaction (TCR) model.
As noted in the literature, while reactions are often discussed in terms of forward and reverse rates, these rates should not be modeled independently because they must be compatible with thermodynamic equilibrium for the system.
Using a simple TCR model where each chemical species has constant heat capacity, we derive the explicit condition that the forward and reverse reaction rate constants must satisfy in order for the system to be thermodynamically consistent.
We perform equilibrium and non-equilibrium simulations of ideal gas mixtures undergoing a reversible dimerization reaction to measure the fluctuational behavior of the system numerically.
We confirm that FHD simulations with the TCR model give the correct static structure factor of equilibrium fluctuations.
For the statistically steady simulation of a gas mixture between two isothermal walls with different temperatures, we show using the TCR model that the temperature variance agrees with the corresponding thermodynamic-equilibrium temperature variance in the interior of the system, whereas noticeable deviations are present in regions near walls, where chemistry is far from equilibrium.
\end{abstract}

\maketitle

\section{\label{sec:Intro}Introduction}

Stochastic modeling of chemical reactions came into prominence in the 1970s.
The chemical master equation~\cite{McQuarrie1967} was formulated based on model birth-death processes, and approximate continuum formulations such as the Fokker--Planck and chemical Langevin equations~\cite{VanKampen2007, Gardiner2004, Gillespie2000} were proposed.
The introduction of Gillespie’s stochastic simulation algorithm (SSA)~\cite{Gillespie1977} further enhanced the popularity.
These stochastic chemistry models were originally limited to simplified reactions (e.g., Schl\"{o}gl model~\cite{Schlogl1972}) and homogeneous systems (e.g., in a continuously stirred tank reactor).

Similar to deterministic reaction-diffusion models where species diffusion and reaction combine, to describe interesting spatio-temporal patterns observed in experiments~\cite{ZhabotinskyZaikin1973}, stochastic chemistry models have been extended to describe spatially inhomogeneous systems.
In the reaction-diffusion master equation (RDME), also known as the multivariate master equation~\cite{Gardiner2004, ErbanChapmanMaini2007}, the system is divided into homogeneous subsystems or cells and diffusive transport of chemical species are described by hopping events into adjacent cells.
Since the direct application of SSA to the RDME is possible in principle but often computationally prohibitive, various simulation techniques have been proposed to accelerate sampling~\cite{ElfEhrenberg2004, WangHouXingYao2011, MarquezLagoBurrage2007, RossinelliBayatiKoumoutsakos2008, IyengarHarrisClancy2010, FuWuLiPetzold2014, KimNonakaBellGarciaDonev2017}.

While these stochastic chemistry models have been successfully employed as theoretical models in various fields of biology~\cite{FangeElf2006, LemarchandNowakowski2011}, the development of such models has been largely divorced from work in the fluid dynamics community.
As a possible reason, we note that such stochastic chemistry models are usually simplified representations aimed at reproducing qualitative features of real chemical systems that may lack some fundamental thermodynamic features.
For example, in the auto-catalytic Brusselator model~\cite{PrigogineLefever1968}, which can explain the spatio-temporal patterns observed in the Belousov-Zhabotinsky reaction, all reactions are irreversible and thus the system cannot reach thermodynamic equilibrium.
By contrast, in the deterministic gas-phase chemical modeling, thermodynamic consistency has been more carefully investigated.
In the resulting mathematical formulation~\cite{Giovangigli2012}, all processes, including chemical reactions, are formulated and analyzed in terms of the chemical potentials of the component species.
The thermodynamic consistency of chemical reactions is established based on the symmetric formulation for the rates of progress.

In this paper, we formulate a thermodynamically-consistent reaction (TCR) model in the form of chemical Langevin equation and construct a thermodynamically-consistent mesoscopic stochastic simulation method for reactive gas mixtures by incorporating the TCR model into a recently-developed fluctuating hydrodynamics (FHD)~\cite{SrivastavaLadigesNonakaGarciaBell2023} model, which provides a stochastic partial differential equation description of mesoscopic fluid systems with thermal noise~\cite{LandauLifschitz1987, ZarateSengers2006, CroccoloZarateSengers2016, GarciaBellNonakaSrivastavaLadigesKim2024}.
The compressible FHD formulation that the authors developed for nonreactive gas mixtures~\cite{BalakrishnanGarciaDonevBell2014,SrivastavaLadigesNonakaGarciaBell2023} is consistent with the aforementioned mathematical formulation for deterministic gas-phase dynamics~\cite{Giovangigli2012}.
Hence, the FHD formulation is based on chemical potentials and guarantees that the fluid system has the correct thermodynamic equilibrium. 
Up to now, however, only simplified, isothermal reactions have been used in reactive FHD~\cite{BhattacharjeeBalakrishnanGarciaAlejandroBellDonev2015, KimNonakaBellGarciaDonev2018}.
To guarantee thermodynamic consistency, we construct the TCR model based on the symmetric formulation for the rates of progress~\cite{Giovangigli2012}.
Assuming that each species has constant heat capacity in the temperature range of interest, we derive closed-form expressions that the reaction rate parameters of the TCR model must satisfy. 

Using the reversible dimerization reaction of $\chem{NO_2}$ as an example, we demonstrate how the reaction rate parameters of the constant-heat-capacity TCR model are determined. 
Using equilibrium and nonequilibrium reactive FHD simulations of $\chem{NO_2}/\chem{N_2O_4}$ gas mixtures, we show how thermodynamic consistency affects the accuracy of mesoscopic stochastic simulations of reactive gas mixtures.
More specifically, we investigate the fluctuating behavior of the system, focusing on the temperature dependence of reaction rate constants and the behavior of resulting temperature fluctuations.

The rest of the paper is organized as follows.
In Section~\ref{sec_RFHD}, we describe the overall structure of our FHD simulation method for ideal gas mixtures undergoing reversible reactions described by the chemical Langevin equation.
In Section~\ref{sec_TCR}, we present the formulation of our TCR model.
We first review general principles and formulate the constant-heat-capacity TCR model.
In Section~\ref{sec_num_example}, we present the results of equilibrium and non-equilibrium simulations of
$\chem{NO_2}/\chem{N_2O_4}$ gas mixtures.
We present our conclusions in Section~\ref{sec_conclusion}.

\section{\label{sec_RFHD}Reactive Fluctuating Hydrodynamics}

In Section~\ref{subsec_overall_struct}, we summarize the overall structure of the reactive fluctuating hydrodynamics formulation~\cite{SrivastavaLadigesNonakaGarciaBell2023} that provides the framework for our thermodynamically-consistent stochastic chemistry formulation.
In Section~\ref{subsec_CLE}, we focus on the reactive source term and discuss how it can be modeled using a chemical Langevin equation (CLE)~\cite{KimNonakaBellGarciaDonev2017, KimNonakaBellGarciaDonev2018}.

\subsection{\label{subsec_overall_struct}Overall Structure}

For a gas mixture of $\Ns$ species, we denote the species mass densities and the total mass density by $\rho_s$ ($s=1,\cdots,\Ns$) and $\rho=\sum_{s=1}^{\Ns} \rho_s$, respectively, and the fluid velocity and the total specific energy (i.e., energy per mass) by $\mathbf{u}$ and $E$, respectively.
The time evolution of the species mass densities ($\rho_s$), momentum density ($\rho\mathbf{u}$), and energy density ($\rho E$) is described by the fluctuating Navier--Stokes (FNS) equations~\cite{BalakrishnanGarciaDonevBell2014, BhattacharjeeBalakrishnanGarciaAlejandroBellDonev2015}:
\begin{subequations}
\label{eq_FNS}
\begin{align}
    \label{eq_FNS_rhos}
    \frac{\partial\rho_s}{\partial t} &=  
    -\nabla\cdot(\rho_s\mathbf{u}) 
    -\nabla\cdot\mathbfcal{F}_s 
    + M_s\Omega_s, \\
    \label{eq_FNS_rhou}
    \frac{\partial(\rho\mathbf{u})}{\partial t} &= 
    -\nabla\cdot(\rho\mathbf{u}\mathbf{u}^T)
    -\nabla{p} 
    - \nabla\cdot\boldsymbol{\Pi}, \\
    \label{eq_FNS_rhoE}
    \frac{\partial(\rho{E})}{\partial t} &= 
    -\nabla\cdot(\rho E\mathbf{u} + p\mathbf{u}) 
    -\nabla\cdot\mathbfcal{Q}
    -\nabla\cdot(\boldsymbol{\Pi}\cdot\mathbf{u}).
\end{align}
\end{subequations}
Here, $M_s \Omega_s$ is the reaction source term where $M_s$ is the molar mass of species $s$ and $\Omega_s$ is the molar concentration production rate of species $s$ due to the chemical reactions associated with species $s$.  
This term will be discussed in more detail in Section~\ref{subsec_CLE}.
The pressure is denoted by $p$, and $\mathbfcal{F}_s$, $\boldsymbol{\Pi}$, and $\mathbfcal{Q}$ are fluxes for the species mass, momentum, and heat, respectively.
We note that $\sum_s \nabla\cdot\mathbfcal{F}_s = 0$ and $\sum_s M_s\Omega_s = 0$.
Consequently, summing Eq.~\eqref{eq_FNS_rhos} gives the equation for conservation of total mass:
\begin{equation}
    \frac{\partial\rho}{\partial t} + \nabla\cdot(\rho\mathbf{u}) = 0.
\end{equation}

While Eqs.~\eqref{eq_FNS} may superficially resemble as the deterministic Navier--Stokes equations, it is important to note that the standard deterministic fluxes for species mass, momentum, and heat are augmented with stochastic components that represent fluctuations.
In other words, these fluxes are expressed as
\begin{equation}
    \mathbfcal{F}_s = \overline{\mathbfcal{F}}_s  + \widetilde{\mathbfcal{F}}_s,\quad
    \boldsymbol{\Pi} = \overline{\boldsymbol{\Pi}} + \widetilde{\boldsymbol{\Pi}},\quad
    \mathbfcal{Q} = \overline{\mathbfcal{Q}} + \widetilde{\mathbfcal{Q}},
\end{equation}
where the overline and tilde notations denote the deterministic and stochastic parts, respectively.
The deterministic fluxes are dissipative and require phenomenological constitutive laws to close the balance equations.
Here, we close the system using the standard phenomenological laws of nonequilibrium thermodynamics, which are expressed in terms of derivatives of species densities, velocity and temperature with transport coefficients that are functions of the thermodynamic state of the fluid.
Note that the parity under time-reversal of the dissipative fluxes differs from the conservative ones, playing an essential role in detailed balance.
The stochastic fluxes are modeled by Gaussian white noise fields with noise intensities that are also functions of the fluid state variables.
For the explicit forms of these expressions, we refer the reader to Refs.~\onlinecite{BalakrishnanGarciaDonevBell2014, SrivastavaLadigesNonakaGarciaBell2023}.

The relation between the total specific energy $E$ and the temperature $T$ is given by
\begin{equation}
    E = \frac{1}{2}|\mathbf{u}|^2 + e(T,\{\rho_s\}).
\end{equation}
Here, the total specific internal energy $e=e(T,\{\rho_s\})$ is a function of temperature and chemical composition.
In this paper, we assume an ideal gas mixture, for which we can simply express $e$ as the weighted sum of the specific internal energy of each species $e_s$:
\begin{equation}
    e(T,\{\rho_s\}) = \frac{1}{\rho} \sum_{s=1}^{\Ns} \rho_s e_s(T).
\end{equation}
Note that the $e_s$ is specified in terms of integrals of specific heats from a reference temperature to $T$.
Data for specific heats, here provided as model inputs, are readily available in both tabular and curve fits.  See, for example, the NIST Chemistry WebBook~\cite{LinstromMallard2001, NISTchemistrywebbook}.
In addition, the equation of state is given by the ideal gas law:
\begin{equation}
    p_s = \frac{\rho_s R T}{M_s},\quad p = \sum_{s=1}^{\Ns} p_s = R T \sum_{s=1}^{\Ns}\frac{\rho_s}{M_s},
\end{equation}
where $p_s$ is the partial pressure of species $s$ and $R$ is the ideal gas constant.

\subsection{\label{subsec_CLE}CLE-based Stochastic Chemistry}

We assume that the chemistry of the gas mixture is described by elementary reactions, which can be grouped into $\Nr$ pairs of reversible reactions.
We write reaction $r$ ($r=1,\cdots,\Nr$) as
\begin{equation}
    \sum_{s=1}^{\Ns} \nu^{+}_{sr} X_s
    \underset{k_r^-}{\stackrel{k_r^+}{\rightleftharpoons}}
    \sum_{s=1}^{\Ns} \nu^{-}_{sr} X_s.
\end{equation}
Here, $X_s$ represents the chemical symbol of species $s$.
We introduce the superscripts $+$ and $-$ to indicate the forward and reverse reactions, respectively, and denote the number of molecules of species $s$ on the reactant side for the forward and reverse reactions as $\nu^+_{sr}$ and $\nu^-_{sr}$, respectively.
We define the stoichiometric coefficients of species $s$ for the reaction $r$ as $\nu_{sr} =\nu_{sr}^{-}-\nu_{sr}^{+}$.
Here, the forward and reverse reaction constants, $k_r^+$ and $k_r^-$, are functions of temperature.

In the chemical Langevin equation (CLE)~\cite{KimNonakaBellGarciaDonev2017, KimNonakaBellGarciaDonev2018}, the molar concentration production rate for species $s$, denoted by $\Omega_s$, is expressed as the sum of deterministic and stochastic parts: 
\begin{equation}
    \Omega_s = \overline{\Omega}_s + \widetilde{\Omega}_s.
\end{equation}
To express $\overline{\Omega}_s$ and $\widetilde{\Omega}_s$, we introduce the mean forward and reverse rates $a_r^+$ and $a_r^-$ of reaction $r$.
Here we assume that the deterministic reaction rates obey the law of mass action for gas phase reactions~\cite{Giovangigli2012, Kondepudi2008}, which expresses $a_r^\pm$ as
\begin{equation}
    a_r^\pm = k_r^\pm \prod_{s=1}^{\Ns} [X_s]^{\nu_{sr}^\pm},
\end{equation}
where $[X_s]$ is the molar concentration of species $s$.
Standard functional forms for the rate constants $k_r^\pm$ and the relationship between these rate constants that is needed to ensure thermodynamic consistency are discussed in Section~\ref{sec_TCR}.

The mean production rate $\overline{\Omega}_s$ is then expressed as
\begin{equation}
    \overline{\Omega}_s = \sum_{r=1}^{\Nr} \nu_{sr} \left( a_r^+ - a_r^- \right).
\end{equation}
Assuming that the occurrences of reactions follow Poisson statistics and applying the Gaussian approximation~\cite{VanKampen2007, Gillespie2000, KimNonakaBellGarciaDonev2018}, one can express the stochastic contribution $\widetilde{\Omega}_s$ as
\begin{equation}
    \widetilde{\Omega}_s = \frac{1}{\sqrt{N_A}}\sum_{r=1}^{\Nr}\nu_{sr} 
    \left( \sqrt{a_r^+}\mathbfcal{W}_r^+ - \sqrt{a_r^-}\mathbfcal{W}_r^- \right),
\end{equation}
where $N_A$ is the Avogadro constant and $\mathbfcal{W}_r^\pm$ denote Gaussian white noise fields satisfying
\begin{equation}
    \langle\mathbfcal{W}_r^\alpha(\mathbf{x},t)\mathbfcal{W}_{r'}^{\alpha'}(\mathbf{x}',t')\rangle
    = \delta_{r,r'}\delta_{\alpha,\alpha'}\delta(\mathbf{x}-\mathbf{x}')\delta(t-t').
\end{equation}
Note that the factor $1/\sqrt{N_A}$ reflects conversion from the variance of the Poisson distribution in terms of number density to molar concentration.
Note also that the Gaussian approximation assumes that the ``cells'' where reactions occur are large enough for composition to be treated as a continuous variable (Gaussian) rather than a discrete one (Poisson).~\cite{VanKampen2007, Gillespie2000}
For the time and spatial discretization of the fluctuating chemistry source term $\widetilde{\Omega}_s$, see Section~S1 in the Supplementary Material.

\section{\label{sec_TCR}Thermodynamically-Consistent Reaction Formulation}

In this section we formulate a thermodynamically-consistent reaction (TCR) model for chemistry.
Although reactions are often discussed in terms of forward and reverse rates, these rates are not independent.
The rates must be compatible with thermodynamic equilibrium for the system.
In Section~\ref{subsec_TCR_principles}, we review the basic thermodynamics and chemical kinetics of ideal gas mixtures and show how thermodynamic equilibrium constrains the forward and reverse rates.
As part of this discussion, we introduce the symmetric form of reactions that arise from kinetic theory as discussed in Ref.~\onlinecite{Giovangigli2012}.
In Section~\ref{subsec_const_heat_cap_TCR_model}, we introduce a constant-heat-capacity TCR model and derive closed-form expressions for the relation between the forward and reverse rate constants.

\subsection{\label{subsec_TCR_principles}Basic Thermodynamics and Chemical Kinetics of Ideal Gas Mixtures}

\subsubsection{\label{subsubsec_TCR_Keq}Chemical Potential and Equilibrium Constant}

The thermodynamics of reaction is based on the chemical potential of the reacting species.
For an ideal gas mixture, the specific chemical potential of each species $s$ is given by~\cite{Giovangigli2012}
\begin{equation}
\label{eq_mus_XT} 
    \mu_{s}([\boldsymbol{X}],T) = \bar{\mu}_{s}^\circ(T) + \frac{R T}{M_s} \log \frac{p_s}{p^{st}}
    = \bar{\mu}_{s}^\circ(T)+\frac{R T}{M_s} \log \frac{R T[X_s]}{p^{st}}
    = {\mu}_{s}^\circ(T)+\frac{R T}{M_s} \log {[X_s]},
\end{equation}
since $p_s = [X_s] R T$.
Here, $[\boldsymbol{X}]$ is a vector containing the molar concentrations $[X_s]$ of each species $s$ and $p^{st}$ is a reference pressure.
Note that $\bar{\mu}_s^\circ$ and ${\mu}_s^\circ = \bar{\mu}_s^\circ + \frac{R T}{M_s}\log \frac{R T}{p^{st}}$ are the chemical potentials of the pure substance at the reference pressure and at unit molar concentration, respectively.
Since we express reaction rates using molar concentrations $[X_s]$ rather than using partial pressures $p_s$ in this paper, we use $\mu_s^\circ$ rather than $\bar{\mu}_s^\circ$. 
However, one can equivalently use $p_s$ and $\bar{\mu}_s^\circ$ and the corresponding results are summarized in Appendix~\ref{appendix_formul_p}.

We use a hat notation to denote a dimensionless per-particle quantity. 
For chemical potential, we define 
\begin{equation}
    \hat{\mu}_s ([\boldsymbol{X}],T) = \frac{M_s}{R T}\mu_s ([\boldsymbol{X}],T),\quad
    \hat{\mu}^\circ_s (T) = \frac{M_s}{R T}\mu^\circ_s (T),
\end{equation}
and thus have
\begin{equation}
\label{eq_muhat_muhatcirc}
    \hat{\mu}_s = \hat{\mu}^\circ_s + \log [X_s].
\end{equation}

For each reaction $r$, the condition $\sum_s \nu_{sr}^+ \hat{\mu}_s = \sum_s \nu_{sr}^- \hat{\mu}_s$ must hold at equilibrium.
Hence, using Eq.~\eqref{eq_muhat_muhatcirc}, at equilibrium we have
\begin{equation}
    \exp\left(- \sum_{s=1}^{\Ns} \nu_{sr} \hat{\mu}_{s}^\circ(T)\right) 
    = \prod_{s=1}^{\Ns} \frac{[X_s]^{\nu_{sr}^-}}{[X_s]^{\nu_{sr}^+}} \;.
\end{equation}
From this equation we can then define the equilibrium constant for reaction $r$ by
\begin{equation}
\label{eq_Keq_chempot}
    K_r(T) = \exp\left(- \sum_{s=1}^{\Ns} \nu_{sr} \hat{\mu}_{s}^\circ(T) \right).
\end{equation}

\subsubsection{\label{subsubsec_TCR_rateconst}Law of Mass Action and Rate Constants}

We note the $K_r(T)$ is a thermodynamic quantity that depends only on the properties of the system at equilibrium; it does not depend on the reaction rates.
In this section, we discuss how $K_r(T)$ constrains the forward and reverse reaction rates.
Note that it is well-established knowledge dating back to Kramers’ theory of chemical reactions.~\cite{Kramers1940, HanggiTalknerBorkovec1990}
Following Ref.~\onlinecite{Giovangigli2012}, we first consider the mean rate of progress of reaction $r$ denoted by $\tau_r = a_r^+ - a_r^-$.
The symmetric form for $\tau_r$, which follows from microscopic reversibility,~\cite{Keizer1987} is given by:
\begin{equation}
    \tau_r = \lambda_r \left\{
    \exp\left(\sum_{s=1}^{\Ns}\nu_{sr}^+ \hat{\mu}_s \right)
    - \exp\left(\sum_{s=1}^{\Ns}\nu_{sr}^-\hat{\mu}_s \right)
    \right\},
\end{equation}
where $\lambda_r$ is the symmetric reaction constant of reaction $r$.
This form of the rate of progress expresses the reaction in terms of a single rate that characterizes how the system relaxes to equilbrium.
Using Eq.~\eqref{eq_muhat_muhatcirc}, we rewrite this equation as
\begin{equation}
    \tau_r = \lambda_r \left\{
    \exp\left(\sum_{s=1}^{\Ns}\nu_{sr}^+\hat{\mu}_s^\circ\right)\prod_{s=1}^{\Ns}[X_s]^{\nu_{sr}^+}
    - \exp\left(\sum_{s=1}^{\Ns}\nu_{sr}^-\hat{\mu}_s^\circ\right)\prod_{s=1}^{\Ns}[X_s]^{\nu_{sr}^-}
    \right\}.
\end{equation}
Hence, by defining the rate constants as 
\begin{equation}
\label{eq_kpm_lambda}
    k^\pm_r(T) = \lambda_r(T) \exp\left(\sum_{s=1}^{\Ns}\nu_{sr}^\pm \hat{\mu}_s^\circ(T)\right),
\end{equation}
we recover the law of mass action:
\begin{equation}
    \tau_r = k_r^+\prod_{s=1}^{\Ns}[X_s]^{\nu_{sr}^+} - k_r^-\prod_{s=1}^{\Ns}[X_s]^{\nu_{sr}^-}.
\end{equation}

At chemical equilibrium, $\tau_r = 0$.
Noting that $k_r^\pm (T)$ share the same factor $\lambda_r (T)$, one can show that for the system to be thermodynamically consistent the forward and reverse rates must be related through the equilibrium constant:
\begin{equation}
\label{eq_Keq_kpm}
    K_r(T) = \frac{k_r^+ (T)}{k_r^- (T)}.   
\end{equation}
Equivalently, one can say that the symmetric constant $\lambda_r (T)$ defines a relation that each pair of rate constants $k_r^\pm (T)$ must satisfy:
\begin{equation}
\label{eq_lambda_cond}
    \lambda_r (T) 
    = k_r^+(T) \exp\left(-\sum_{s=1}^{\Ns}\nu_{sr}^+\hat\mu_s^\circ(T)\right)
    = k_r^-(T) \exp\left(-\sum_{s=1}^{\Ns}\nu_{sr}^-\hat\mu_s^\circ(T)\right).
\end{equation}
The relationship~\eqref{eq_Keq_kpm} is required for thermodynamic consistency.
Consequently, for thermodynamic consistency only one of the forward and reverse rate constants can be specified; the other must be obtained via Eq.~\eqref{eq_Keq_kpm} or equivalently Eq.~\eqref{eq_lambda_cond}.

In practice, the functional form of a reaction rate constant is modeled by an empirical expression.
The most commonly used form is the modified Arrhenius equation~\cite{Laidler1996, IUPAC1997}:
\begin{equation}
    k(T) = \mathfrak{A}\; \left(\frac{T}{T^{st}}\right)^\beta e^{-\alpha/R T},
\end{equation}
where $\mathfrak{A}$ is a temperature-independent constant, $\beta$ is a dimensionless number, $\alpha$ has units of energy per mole, and $T^{st}$ denotes a standard or reference temperature. 
While some theoretical justifications are available for this form~\cite{Kondepudi2008}, its parameters are usually determined by fitting reaction rate data to this form.
Note that this form is used in several standard chemical kinetics databases (e.g.\ the NIST chemical kinetics database~\cite{NISTchemkin} or CHEMKIN~\cite{KeeRupleyMeeksMiller1996}) and the parameter values are available. 

We note that even if one of the rate constants, say $k^+(T)$, is modeled by the modified Arrhenius equation, the other rate constant, $k^-(T)$, will not, in general, be described by the same form.
This is because $k^-(T)$ cannot be specified independently.

\subsection{\label{subsec_const_heat_cap_TCR_model}Constant-Heat-Capacity TCR Model}

We now consider a simplified TCR model that nevertheless still enables realistic modeling of gas-phase reactions.
We show that, in the resulting chemistry formulation with the modified Arrhenius equation, the general condition required for thermodynamic consistency, see Eqs.~\eqref{eq_Keq_kpm} or \eqref{eq_lambda_cond}, can be reduced to simple relations between rate-constant parameters, see Eqs.~\eqref{eq_KT_KTst}, \eqref{eq_kpmT_kpmTst}, and \eqref{eq_CHC_TCR_param_rel}. 
We use this model in the numerical example described in Section~\ref{sec_num_example}.

\subsubsection{\label{subsubsec_CHC_TCR_assumption}Assumption}

We assume that the specific heat at constant pressure, $c_{p,s}$, is constant for each species $s$.
The specific enthalpy of species $s$ can then be expressed as a linear function of temperature,
\begin{equation}
\label{eq_hsT}
    h_s(T) = h_s^{st} + \int_{T^{st}}^T c_{p,s}\: dT' = h_s^{st} + c_{p,s}(T-T^{st}).
\end{equation}
Since the specific internal energy $e_s(T)$ of species $s$ is given as $e_s(T) = h_s(T) - R T / M_s$, it can also be expressed as a linear function of temperature where the specific heat at constant volume, $c_{v,s}$, is also constant so that
\begin{equation}
\label{eq_esT}
    e_s(T) = e_s^{st} + c_{v,s}(T-T^{st}),
\end{equation}
where
\begin{equation}
    e_s^{st} = h_s^{st} - \frac{R T^{st}}{M_s},\quad 
    c_{v,s} = c_{p,s} - \frac{R}{M_s}.
\end{equation}
One can also derive the pure-component specific entropy of species $s$:
\begin{equation}
\label{eq_ssT}
    s^\circ_s(T) = s_s^{st} + \int_{T^{st}}^T \frac{c_{p,s}}{T'} dT' = s_s^{st} + c_{p,s} \log \frac{T}{T^{st}}.
\end{equation}

We note that the values of $c_{p,s}$, $h_s^{st}$, and $s_s^{st}$ are readily available in thermochemical databases (e.g., the NIST Chemistry WebBook~\cite{LinstromMallard2001, NISTchemistrywebbook}).
Although the specific heat capacity is actually a function of temperature, the constant specific heat capacity model remains valid in a rather wide temperature range in many cases as demonstrated using the $\mathrm{NO_2}$ dimerization example in Section~\ref{sec_num_example}.

\subsubsection{\label{subsubsec_CHC_TCR_relations}Equilibrium Constant and Rate Constants}

We show here that for our constant-heat-capacity TCR model the general expression for $K_r(T)$, Eq.~\eqref{eq_Keq_chempot}, can be written in the form
\begin{equation}
    K(T)\sim \left(\frac{T}{T^{st}}\right)^B e^{-A/R T}
\end{equation}
for some choice of parameters $A$ and $B$, which is identical to the modified Arrhenius equation for rate constants.
As discussed above, we can write the chemical potential in terms of enthalpy and entropy,
\begin{equation}
\label{eq_muhat_h_s}
    \hat\mu_s^\circ (T) = \frac{M_s}{R T} \{h_s(T)-T s^\circ_s(T) \} 
    + \log \frac{RT}{P^{st}}, \quad
    \hat\mu_s^\circ (T^{st}) = \frac{M_s}{R T^{st}} (h_s^{st} - T^{st} s_s^{st})
    + \log \frac{RT^{st}}{P^{st}}.
\end{equation}
Note that logarithm terms appear because we use $\mu_s^\circ$ instead of $\overline{\mu}_s^\circ$, see Eq.~\eqref{eq_mus_XT}.
For the simplified model considered here, from Eqs.~\eqref{eq_hsT} and \eqref{eq_ssT}, we obtain
\begin{equation}
\label{eq_muhatT_muhatTst_intermediate}
    \hat{\mu}_s^\circ(T) - \hat{\mu}_s^\circ(T^{st}) 
    = \frac{M_s (h_s^{st} - c_{p,s} T^{st})}{R}\left(\frac{1}{T}-\frac{1}{T^{st}}\right)
    - \left(\frac{M_s}{R} c_{p,s} -1 \right) \log\frac{T}{T^{st}}.
\end{equation}

By defining the internal energy (or enthalpy) extrapolated to zero temperature,
\begin{equation}
\label{eq_eps_def}
    \epsilon_s = h_s^{st} - c_{p,s} T^{st} = e_s^{st} - c_{v,s} T^{st},
\end{equation}
and the unitless per-molecule specific heat at constant volume,
\begin{equation}
\label{eq_cvhat_def}
    \hat{c}_{v,s} = \frac{M_s}{R}c_{v,s} = \frac{M_s}{R}\left(c_{p,s} - \frac{R}{M_s}\right),
\end{equation}
we rewrite Eq.~\eqref{eq_muhatT_muhatTst_intermediate} as
\begin{equation}
\label{eq_muhatT_muhatTst}
    \hat{\mu}_s^\circ(T) - \hat{\mu}_s^\circ(T^{st}) 
    = \frac{M_s \epsilon_s}{R}\left(\frac{1}{T}-\frac{1}{T^{st}}\right)
    - \hat{c}_{v,s} \log \frac{T}{T^{st}}.
\end{equation}
By substituting Eq.~\eqref{eq_muhatT_muhatTst} into Eq.~\eqref{eq_Keq_chempot}, we finally obtain
\begin{equation}
\label{eq_KT_KTst}
    K_r (T) = K_r (T^{st}) \exp\left[-\frac{A_r}{R}\left(\frac{1}{T}-\frac{1}{T^{st}}\right)\right]
    \left(\frac{T}{T^{st}}\right)^{B_r},
\end{equation}
where
\begin{equation}
\label{eq_Ar_Br_def}
    A_r = \sum_{s=1}^{\Ns} \nu_{sr} M_s\epsilon_s,\quad
    B_r = \sum_{s=1}^{\Ns} \nu_{sr} \hat{c}_{v,s}.
\end{equation}

The form of $K_r(T)$ given by Eq.~\eqref{eq_KT_KTst} for this simplified TCR model implies that, if one of the rate constants can be written in the modified Arrhenius form, then the other can be as well. 
In particular, when we write the forward and reverse rate constants in the modified Arrhenius form as 
\begin{equation}
\label{eq_kpmT_kpmTst}
    k_r^\pm (T) = k_r^\pm (T^{st}) \exp\left[-\frac{\alpha_r^\pm}{R}\left(\frac{1}{T}-\frac{1}{T^{st}}\right)\right]
    \left(\frac{T}{T^{st}}\right)^{\beta_r^\pm},
\end{equation}
using relation~\eqref{eq_Keq_kpm}, the relation between forward and reverse rates can then be simplified to 
\begin{equation}
\label{eq_CHC_TCR_param_rel}
    \alpha_r^- = \alpha_r^+ - A_r, \quad
    \beta_r^- = \beta_r^+ - B_r.
\end{equation}

As an aside, we see that if the forward and reverse rate constants, $k^\pm_r$, are \textit{independent} of temperature then thermodynamic consistency requires that
\begin{equation}
    \sum_{s=1}^{\Ns} \nu_{sr}^+ M_s\epsilon_s  
    =  \sum_{s=1}^{\Ns} \nu_{sr}^- M_s\epsilon_s,\quad
    \sum_{s=1}^{\Ns} \nu_{sr}^+ \hat{c}_{v,s}  =  
    \sum_{s=1}^{\Ns} \nu_{sr}^- \hat{c}_{v,s},
\end{equation}
for all reactions.
For example, in a reversible dimerization reaction, $\chem{2A}{\rightleftharpoons}\chem{A_2}$, these conditions become $2 M_1 \epsilon_1 = M_2 \epsilon_2$ and $2 \hat{c}_{v,1} = \hat{c}_{v,2}$, where species 1 and 2 are the monomer and dimer, respectively.
Since $2M_1=M_2$, the former condition is further reduced to $\epsilon_1 = \epsilon_2$.

When the TCR model is incorporated into a mesoscopic hydrodynamic description, it is important to use the instantaneous temperature to evaluate reaction rate constants $k_r^\pm(T)$ for fluctuations for the overall system to be correct.
In the following section, using a dimerization reaction, we show numerically that this is an essential requirement of the TCR model.
In Appendix~\ref{appendix_structfac}, we also provide a theoretical justification by analyzing the fluctuational behavior of the equilibrium system via the static structure factor spectra.

\section{\label{sec_num_example}Numerical Example}

In this section, we consider an ideal gas mixture of $\chem{NO_2}$ and $\chem{N_2O_4}$ undergoing a reversible dimerization reaction,
\begin{equation}
\label{eq_dimerization_reaction}
    \chem{2NO_2}\underset{k^-}{\stackrel{k^+}{\rightleftharpoons}}\chem{N_2O_4},
\end{equation}
and perform reactive FHD simulations.
In Section~\ref{subsec_reaction_params}, we describe how to model the reaction and determine reaction parameters.
In Section~\ref{subsec_RFHD_implementation}, we briefly explain our numerical implementation of reactive FHD and the determination of the FHD simulation parameters.
In Section~\ref{subsec_eq_sim_res}, by performing equilibrium simulations and confirming that the correct equilibrium distribution is established, we validate our theoretical formulation and numerical implementation. 
In Section~\ref{subsec_neq_sim_res}, we consider a nonequilibrium steady state maintained between two thermal walls with different temperatures and study nontrivial nonequilibrium fluctuational behavior of the system due to the reaction.

\subsection{\label{subsec_reaction_params}Reaction Modeling and Parameters}

Our simulation model is inspired by the experimental study of the dimerization of nitrogen dioxide~\cite{BorrellCobosLuther1988}. 
While the chemical kinetics was measured in the presence of nitrogen gas in the study, we first consider the case where the ideal gas mixture only contains $\chem{NO_2}$ and $\chem{N_2O_4}$ and then investigate the case where (non-reactive) $\chem{N_2}$ is also included.
For the species index $s$, $\chem{NO_2}$ is set to $s=1$ and $\chem{N_2O_4}$ to $s=2$.
When nitrogen is added, $\chem{N_2}$ is set to $s=3$.
Since we only consider one reversible reaction (i.e., $\Nr=1$), we drop the reaction index $r$.
By the law of mass action, the forward and reverse rates are expressed as $a^+ = k^+[\chem{NO_2}]^2$ and $a^-=k^-[\chem{N_2O_4}]$, respectively.
The chemical production rates of the species are given as
\begin{align}
    \Omega_1 &= -2\left(a^+ + \sqrt{\frac{a^+}{N_A}}\mathbfcal{W}^+\right) 
    +2 \left( a^- + \sqrt{\frac{a^-}{N_A}}\mathbfcal{W}^-\right),\\
    \Omega_2 &= \left(a^+ + \sqrt{\frac{a^+}{N_A}}\mathbfcal{W}^+ \right) 
    - \left( a^- + \sqrt{\frac{a^-}{N_A}}\mathbfcal{W}^- \right).
\end{align}
We use cgs units and, particularly, \SI{}{\mol/\cm^3} for molar concentration so that the units of $k^+$ and $k^-$ are \SI{}{\cm^3 /\mol\cdot\s} and \SI{}{\s^{-1}}, respectively.

To model the forward and backward reaction rate constants, we use the constant-heat-capacity TCR model formulated in Section~\ref{subsec_const_heat_cap_TCR_model}.
We thus assume that each chemical species has constant heat capacity and the reaction rate constants follow the modified Arrhenius equation.
The equilibrium constant $K(T)$ is then given by Eq.~\eqref{eq_KT_KTst} with parameters $K(T^{st})$, $A$, and $B$.
Using the thermochemistry data of $\chem{NO_2}$ and $\chem{N_2O_4}$ in the NIST Chemistry WebBook~\cite{LinstromMallard2001}, we determine the values of these parameters as follows.
For each species, we first compute the values of $h_s^{st}$, $\s_s^{st}$, and $c_{p,s}$ by evaluating the Shomate equation at $T^{st}=\SI{350}{\K}$ and $p^{st}=10^6\:\mathrm{Ba}$.
By using Eqs.~\eqref{eq_muhat_h_s}, \eqref{eq_eps_def}, and \eqref{eq_cvhat_def}, we determine the values of $\hat{\mu}^\circ_s(T^{st})$, $\epsilon_s$, and $\hat{c}_{v,s}$, which are given in Table~\ref{table_thermochem_params}.
By using Eqs.~\eqref{eq_Keq_chempot} and \eqref{eq_Ar_Br_def}, we finally obtain $K(T^{st}) = \mbox{\SI{6.52e3}{}}$, $A = \mbox{\SI{-5.90e11}{\erg/\mol}}$, and $B = \mbox{\SI{1.74}{}}$.
We obtain the parameter values of the reverse reaction, $k^-(T^{st})$, $\alpha^-$, and $\beta^-$, from Ref.~\citenum{BorrellCobosLuther1988} via NIST Chemical Kinetics Database~\cite{NISTchemkin}.
We then determine the parameter values of the forward reaction, $k^+(T^{st})$, $\alpha^+$, and $\beta^+$, using Eqs.~\eqref{eq_Keq_kpm} and \eqref{eq_CHC_TCR_param_rel}.
The values of the reaction parameters, $k^\pm(T^{st})$, $\alpha^\pm$, and $\beta^\pm$, are given in Table~\ref{table_react_params}.

\begin{table}
\begin{ruledtabular}
\begin{tabular}{lccc}
& units & $\chem{NO_2}$ ($s=1$) & $\chem{N_2O_4}$ ($s=4$) \\
\hline
$\hat{\mu}^\circ_s(T^{st})$ & unitless      & \SI{-7.27}{}   & \SI{-23.3}{}   \\
$\varepsilon_s$ & \SI{}{\erg/\g}            & \SI{4.69e9}{}  & \SI{-1.73e9}{} \\
$\hat{c}_{v,s}$ & unitless                  & \SI{3.64}{}    & \SI{9.02}{}    \\
\end{tabular}
\end{ruledtabular}
\caption{\label{table_thermochem_params}
Thermodynamic parameter values for $\chem{NO_2}$ and $\chem{N_2O_4}$.
These values are obtained for $T^{st}=\SI{350}{\K}$ and $p^{st}=10^6\:\mathrm{Ba}$ using the thermochemistry data of $\chem{NO_2}$ and $\chem{N_2O_4}$ in the NIST Chemistry WebBook~\cite{LinstromMallard2001}.}
\end{table}

\begin{table}
\begin{ruledtabular}
\begin{tabular}{lccc}
& units & forward reaction ($+$) & reverse reaction ($-$) \\
\hline
$k^\pm(T^{st})$ & $+$: \SI{}{\cm^3 /\mol\cdot\s}, $-$: \SI{}{\s^{-1}}  & \SI{4.07e11}{}     & \SI{6.24e07}{} \\
$\alpha^\pm$    & \SI{}{\erg/\mol}          & \SI{-5.34e10}{}    & \SI{5.37e11}{} \\
$\beta^\pm$     & unitless                  & \SI{0.645}{}       & \SI{-1.10}{}   \\
\end{tabular}
\end{ruledtabular}
\caption{\label{table_react_params}
Reaction parameter values for the dimerization reaction~\eqref{eq_dimerization_reaction}.
These parameters appear in Eq.~\eqref{eq_kpmT_kpmTst}}
\end{table}

\begin{figure}
    \centering
    \includegraphics[width=\linewidth]{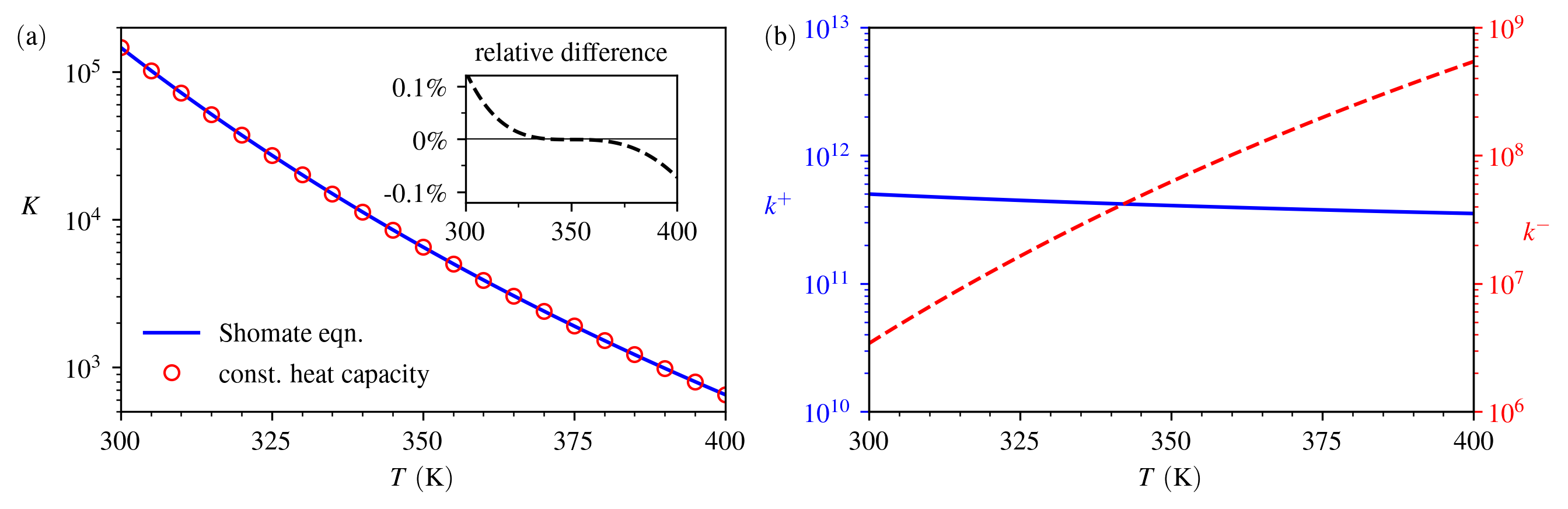}
    \caption{\label{fig:K_kp_km}
    In panel~(a), the values of the equilibrium constant $K(T)$ obtained from our constant-heat-capacity TCR model are compared with those computed by the Shomate equation in the temperature range $\mbox{\SI{300}{\K}}\le T\le\mbox{\SI{400}{\K}}$.
    Their relative differences, $(K_\mathrm{model}-K_\mathrm{Shomate})/K_\mathrm{Shomate}$, are shown in the inset. 
    In panel~(b), the values of the forward and reverse reaction rate constants, $k^+ (T)$ and $k^-(T)$, are plotted versus $T$.
    Two $y$-axes are used; the left one is for $k^+(T)$ (depicted by the blue solid line), whereas the right one is for $k^-(T)$ (red dashed line).}
\end{figure}

Figure~\ref{fig:K_kp_km} shows the values of $K(T)$, $k^+(T)$, and $k^-(T)$ computed from our constant-heat-capacity TCR model.
In panel~(a), we compare the equilibrium constant values of our model with those directly obtained at each temperature by evaluating the Shomate equation\cite{LinstromMallard2001}, where the temperature dependence of heat capacity is considered.
It is remarkable that our constant-heat-capacity TCR model reproduces the equilibrium constant $K(T)$ faithfully (within about 0.1\% errors) in the temperature range $\mbox{\SI{300}{\K}}\le T\le\mbox{\SI{400}{\K}}$, while the value of $K(T)$ significantly changes from $K(T^{st})$. 
In panel~(b), the temperature dependence of $k^+(T)$ and $k^-(T)$ is compared.
As temperature increases, $k^+(T)$ decreases gradually, whereas $k^-(T)$ increases significantly.  
These behaviors can be mainly explained by the signs and magnitudes of the parameters $\alpha^+$ and $\alpha^-$.
We note that the value of $\alpha^+$, which is computed from $\alpha^-$ and $A$ via Eq.~\eqref{eq_CHC_TCR_param_rel}, is negative but rather small.

\subsection{\label{subsec_RFHD_implementation}Implementation of Reactive FHD}

We construct a numerical scheme to solve the reactive FNS equations~\eqref{eq_FNS} by incorporating the chemistry source terms $M_s\Omega_s$ into the existing (nonreactive) FNS solver~\cite{SrivastavaLadigesNonakaGarciaBell2023}.
We summarize the main features of our numerical method here and refer the reader to Section~S1 in the Supplementary Material for a detailed description of our numerical implementation of reactive FHD.
The FNS solver~\cite{SrivastavaLadigesNonakaGarciaBell2023} is based on a method-of-lines approach~\cite{DonevVandenEijndenGarciaBell2010}.
In other words, the FNS equations are discretized first in space and the resulting stochastic ordinary differential equations are solved by a stochastic time integration scheme.
For spatial discretization, a staggered spatial discretization~\cite{BalboaBellDelgadoBuscalioniDonevFaiGriffithPeskin2012} is used so that species mass densities $\rho_s$ and energy density $\rho E$ are located at cell centers, whereas momentum density $\rho\mathbf{u}$ is at cell faces.
For temporal integration, an explicit, three-stage, low-storage Runge--Kutta (RK3) scheme~\cite{DonevVandenEijndenGarciaBell2010} is employed.
By assuming that chemistry is not stiff, we include chemistry as source terms in the RK3 scheme.
Since chemistry source terms are given in the form of Langevin equations, they are readily incorporated into the RK3 scheme.
Note that, since $\rho_s$ and $\rho E$ (and thus temperature $T$) are located at cell centers, the chemistry solver can also be incorporated into a fully cell-centered FNS solver~\cite{BalakrishnanGarciaDonevBell2014}.

We determine the FHD simulation parameters as follows.
The reference equilibrium state is chosen so that the sum of partial pressures of $\chem{NO_2}$ and $\chem{N_2O_4}$ is equal to \SI{1}{\atm} at $T^{eq}=T^{st}=\mbox{\SI{350}{\K}}$.
Equilibrium species mass densities can be computed at chemical equilibrium to obtain: $\rho_1^{eq}=\mbox{\SI{1.35e-3}{\g/\cm^3}}$ and $\rho_2^{eq}=\mbox{\SI{5.13e-4}{\g/\cm^3}}$.
The simulation box is a cube of side length $L=\mbox{\SI{2.56e-4}{\cm}}$ and is discretized into $32^3$ cells of side length $\Delta x =\mbox{\SI{8e-6}{\cm}}$.
For equilibrium simulations in Section~\ref{subsec_eq_sim_res}, periodic boundary conditions are used for all variables.
For nonequilibrium simulations in Section~\ref{subsec_neq_sim_res}, physical boundaries with two parallel thermal walls are considered.
The corresponding boundary conditions are described in Section~\ref{subsec_neq_sim_res}.
We use time integration step size $\Delta{t}=10^{-12}\:\mbox{\SI{}{\s}}$.
Note that thermochemistry parameters, $\epsilon_s$ and $\hat{c}_{v,s}$, which are used to determine the reaction parameters, are also used to define the internal energy of each species, $e_s(T) = \epsilon_s + \frac{R}{M_s}\hat{c}_{v,s}T$.
For \chem{N_2}, $\epsilon_3=\mbox{\SI{-3.10e9}{\erg/\g}}$ and $\hat{c}_{v,3}=2.51$.
Finally, since our focus is on the static properties of thermodynamic fluctuations (e.g., structure factor), for computational efficiency we assume the hydrodynamics to be that of a hard-sphere gas mixture.
To evaluate transport coefficients~\cite{HirschfelderCurtissBird1954} (e.g., viscosity), the species are assumed to be hard spheres with diameters of $d_1=\mbox{\SI{3.8e-8}{\cm}}$, $d_2=\mbox{\SI{4.6e-8}{\cm}}$, and $d_3=\mbox{\SI{3.0e-8}{\cm}}$.

\subsection{\label{subsec_eq_sim_res}Equilibrium Simulations}

To demonstrate that reactive FHD based on our thermodynamically-consistent chemistry formulation reproduces the correct statistical distributions at equilibrium, we first perform equilibrium simulations of the $\chem{NO_2}/\chem{N_2O_4}$ mixture.
To this end, one could compute equilibrium statistics in real space by measuring the covariance matrix. 
However, we here investigate the static structure factor~\cite{DonevVandenEijndenGarciaBell2010}.
While the covariance matrix and the structure factor contain essentially the same information, the latter is more convenient to compute and easier to visualize and analyze.
The structure factor for the field variable $\phi$ is defined as
\begin{equation}
\label{eq_struct_fac_def}
    S_{\phi}(\mathbf{k}) = \Delta V\langle{\delta\widehat{\phi}_\mathbf{k}}{\delta\widehat{\phi}_\mathbf{k}^{\ast}}\rangle,
\end{equation}
where brackets $\langle\cdot\rangle$ denote equilibrium average, $\Delta V$ is the volume of a FHD cell, $\mathbf{k}=(k_x,k_y,k_z)$ is a vector of wavenumbers, $\delta\widehat{\phi}_\mathbf{k}$ is the discrete Fourier transform coefficient of $\delta\phi=\phi-\phi^{eq}$ for wavevector $\mathbf{k}$, and $\delta\widehat{\phi}_\mathbf{k}^{\ast}$ is its complex conjugate.
To estimate the equilibrium average in Eq.~\eqref{eq_struct_fac_def}, we run each simulation up to $2\times10^6$ time steps and compute the time average with the first $2\times10^5$ time steps discarded.

When thermodynamic equilibrium is reached (i.e., exhibiting the correct cell variance value with zero covariance values across different cells), the structure factor of each field becomes a constant function (i.e., $S_\phi(\mathbf{k})=S_{\phi,eq}$) and its constant value $S_{\phi,eq}$ is related to the correct cell variance via 
\begin{equation}
\label{eq_Sk_cellvar_rel}
    S_{\phi,eq}=\Delta V\langle\delta\phi^2\rangle.   
\end{equation}
For example, since the equilibrium cell variance of temperature is given as
\begin{equation}
\label{eq_Tvar}
    \langle{\delta{T}^2}\rangle = \frac{\kb (T^{eq})^2}{{\rho^{eq}}c_{v,mix}\Delta V},
\end{equation}
where $\kb$ is the Boltzmann constant, $\rho^{eq}=\sum_{s=1}^{\Ns}\rho^{eq}_s$, and $c_{v, mix} = \frac{1}{\rho^{eq}}\sum_{s=1}^{\Ns} \rho_s^{eq} c_{v,s}$, the structure factor of the temperature field is given as
\begin{equation}
    S_T(\mathbf{k}) = S_{T,eq} = \frac{\kb (T^{eq})^2}{{\rho^{eq}}c_{v,mix}}.
\end{equation}

\begin{figure}
    \centering
    \includegraphics[width=\linewidth]{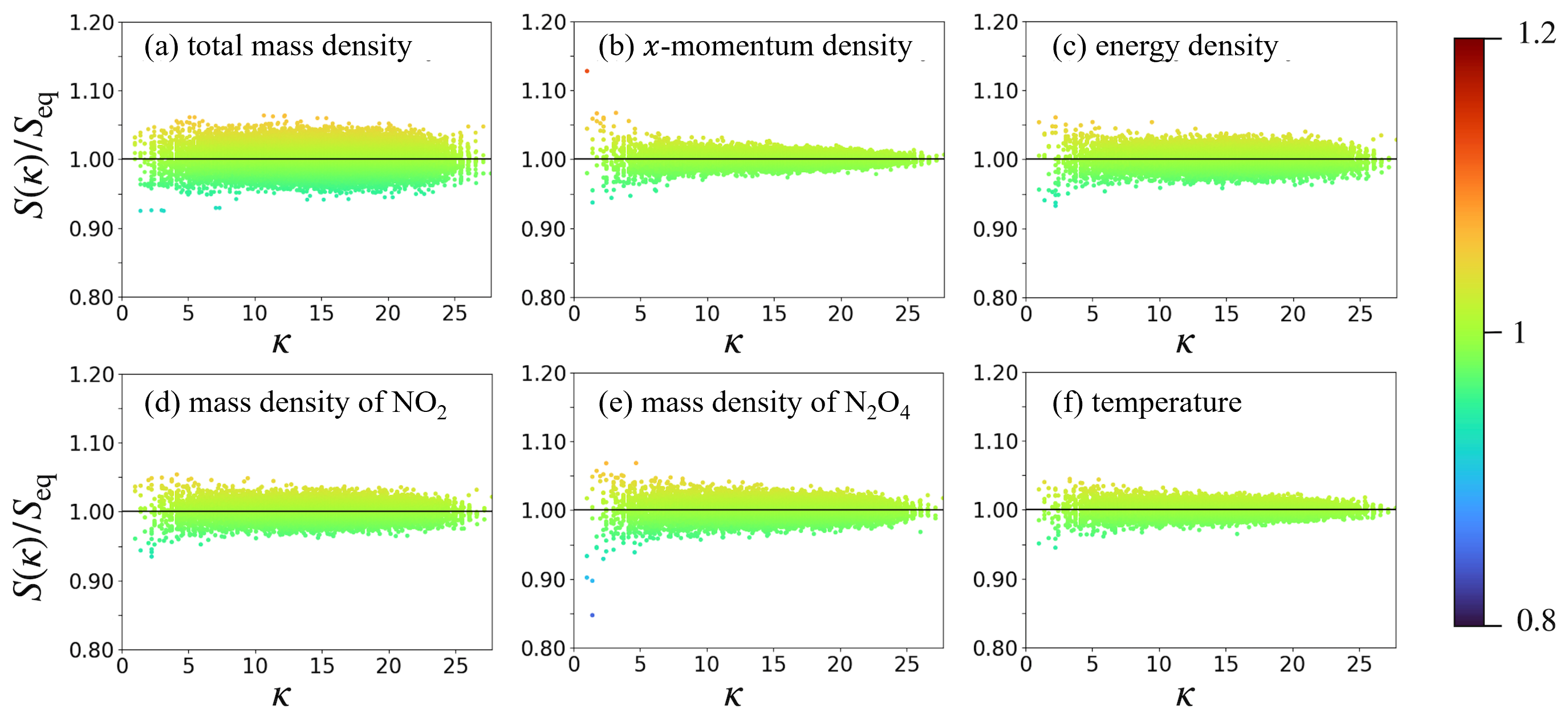}
    \caption{\label{fig:sr_all_fields}
    Equilibrium structure factor spectra obtained from reactive FHD based on our thermodynamically-consistent chemistry formulation.
    For various field variables, the structure factor values $S(\mathbf{k})$ are normalized by the theoretical values $S_{eq}$ (see the color bar) and plotted versus $\kappa=\sqrt{\kappa_x^2 + \kappa_y^2 + \kappa_z^2}$, where $\kappa_i = k_i \left(2\pi/L\right)^{-1}$ is the integer wave index in the $i$-direction.
    Panels show scatter plots for the structure factors of (a) $\rho$, (b) $\rho u_x$, (c) $\rho E$, (d) $\rho_1$, (e) $\rho_2$, and (f) $T$.
    The horizontal black lines at unity show the theoretically expected value.}
\end{figure}

For various field variables ($\rho$, $\rho u_x$, $\rho E$, $\rho_1$, $\rho_2$, and $T$), we show in Figure~\ref{fig:sr_all_fields} the static structure factors that we obtain from our equilibrium simulation.
For each field $\phi$, we normalize the values of $S_\phi(\mathbf{k})$ by the theoretical structure factor value $S_{\phi,eq}$, which can be computed from the theoretical equilibrium cell variance~\cite{BalakrishnanGarciaDonevBell2014} via Eq.~\eqref{eq_Sk_cellvar_rel}.
To see whether $S_\phi(\mathbf{k})$ does not depend on $\mathbf{k}$, we draw a scatter plot for the normalized structure factor values $S_{\phi}(\mathbf{k})/S_{\phi,eq}$ versus $\kappa=\sqrt{\kappa_x^2+\kappa_y^2+\kappa_z^2}$, where $\kappa_i = k_i \left(2\pi/L\right)^{-1}$ is the integer wave index in the $i$-direction.
We observe that all structure factors exhibit flat spectra with the correct values.
For the corresponding nonreactive system (i.e., with $k^+=k^-=0$), we observe the same behavior (see Figure~S1 in the Supplementary Material).
Hence, these simulation results confirm that the incorporation of our stochastic formulation does not disturb the thermodynamic equilibrium state correctly established by nonreactive FHD.
In Appendix~\ref{appendix_structfac}, we also analytically show that our thermodynamically-consistent chemistry formulation gives flat structure factors with the correct values by analyzing the chemistry source terms $M_s\Omega_s$ in the FNS equations~\eqref{eq_FNS} and the structure of the linearized FNS equations for dimerization.

\begin{figure}
    \centering
    \includegraphics[width=\linewidth]{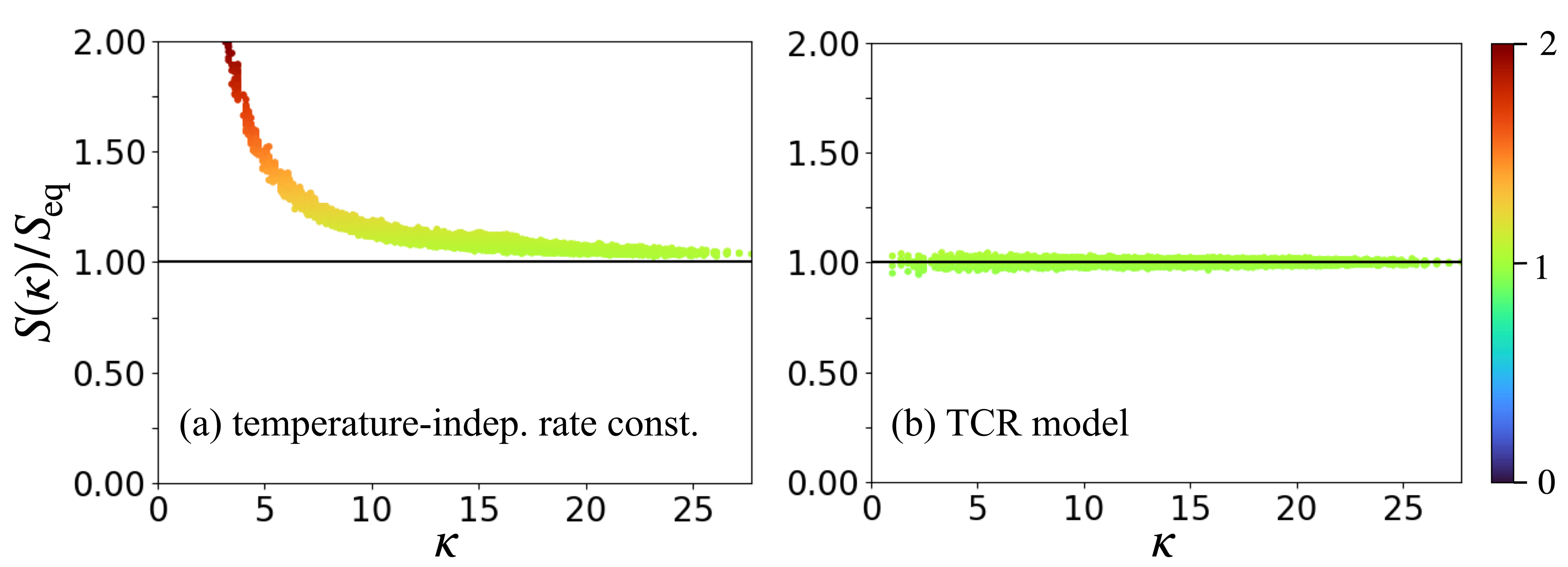}
    \caption{\label{fig:temp_Sr}
    Temperature equilibrium structure factor spectra obtained from (a)~the temperature-independent rate constants case and (b)~our TCR model are compared.
    Note that the same data are shown in panel~(b) of this figure and panel~(f) of Figure~\ref{fig:sr_all_fields} but different vertical scales are used.}
\end{figure}

We now turn into the case where a thermodynamically-inconsistent chemistry formulation is used.
In an equilibrium simulation, particularly with small temperature fluctuations (note that the magnitude of temperature fluctuations in our dimerization example is $\langle\delta T^2\rangle^{1/2}=\mbox{\SI{1.6}{\K}}$), one is tempted to use temperature-independent rate constants, i.e., $k^\pm(T) = k^\pm(T^{eq})$.
Specifically, using temperature-independent rate constants refers to replacing the dependence on fluctuating temperature with the constant average equilibrium temperature.
However, when these rates are used in FHD simulations, remarkably significant errors appear in the structure factors.
Figure~\ref{fig:temp_Sr} compares the temperature structure factor of the temperature-independent rate constants case with that of our TCR model.
In the former case, the structure factor spectrum is not flat and more significant deviations are observed for smaller wave vectors $\mathbf{k}$.
This implies that temperature fluctuations in different fluid cells are correlated and these incorrect correlations are long-ranged.
On the other hand, at short length scales (i.e., large $\mathbf{k}$), fluctuation behavior is similar to those of the nonreactive FHD system. 
As shown in Figure~S2 in the Supplementary Material, the structure factor spectra of the other field variables exhibit similar patterns.
In Appendix~\ref{appendix_structfac}, we provide an explanation of why thermodynamically-inconsistent chemistry causes errors in the small $\mathbf{k}$ region of the structure factor spectra.
Roughly speaking, the impact of fluctuations arising from reaction becomes dominant at long-length scales, whereas that from transport processes (e.g.\ diffusion) becomes dominant at short-length scales. 
For a stochastic reaction-diffusion system, a similar behavior was discussed with the concept of the penetration depth~\cite{KimNonakaBellGarciaDonev2017}.

\begin{figure}
    \centering
    \includegraphics[width=0.5\linewidth]{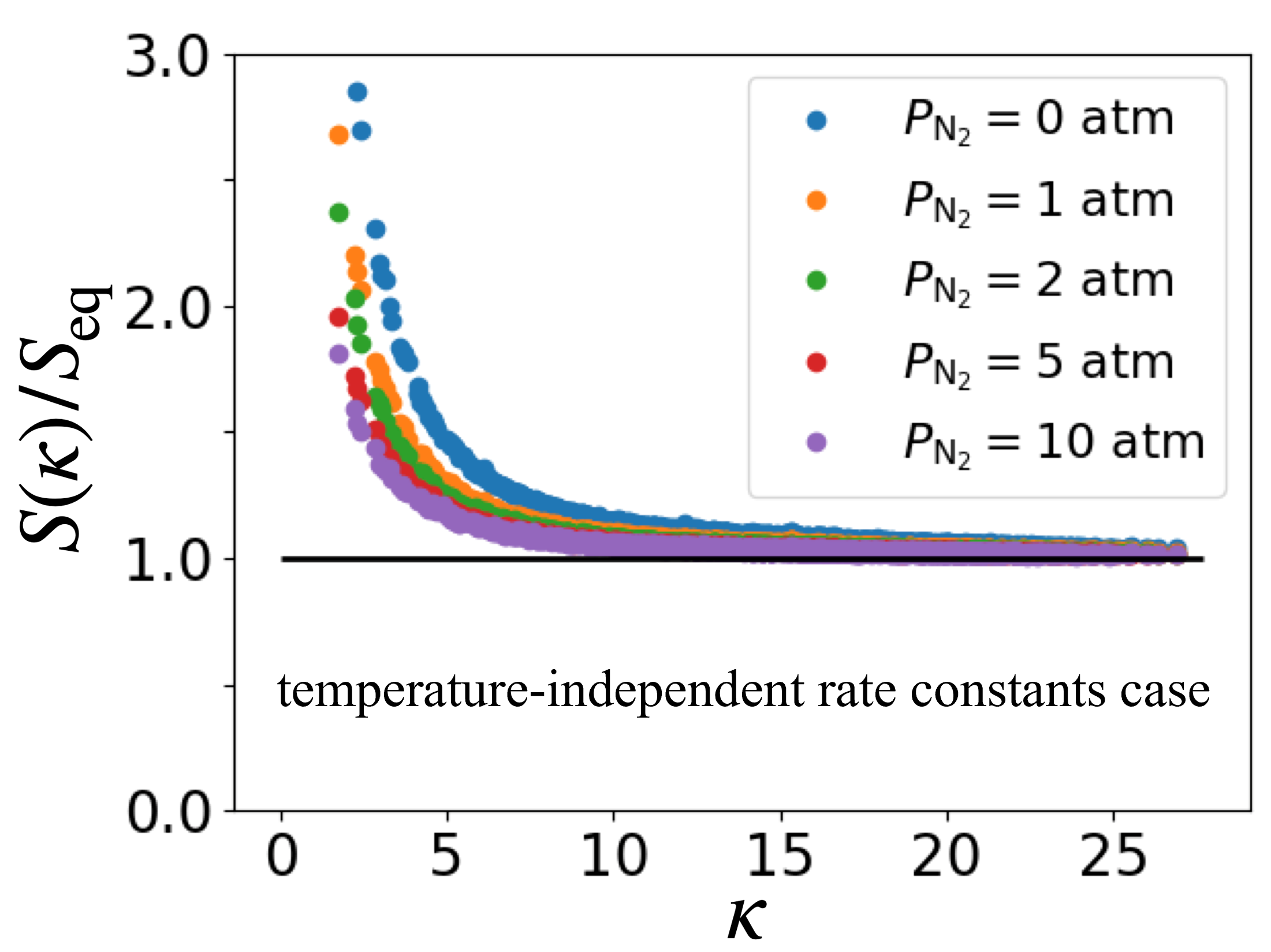}
    \caption{\label{fig_N2sim}
    For various partial pressure values of $\chem{N_2}$, the temperature structure factor spectra obtained from the temperature-independent rate constants case are compared.
    For visual clarity, rather than plotting all data points, representative values of $S(\kappa)/S_{eq}$ obtained by averaging within each subinterval of $\kappa$  are shown.}
\end{figure}

We also perform reactive FHD simulations with molecular nitrogen added to the mixture and compare the structure factor spectra of our TCR model and the temperature-independent rate constants case. 
While maintaining the sum of the equilibrium partial pressures of $\chem{NO_2}$ and $\chem{N_2O_4}$ to be \SI{1}{\atm}, we consider ideal gas mixtures with $P_\chem{N_2} = 1$, 2, 5, and \SI{10}{\atm}. 
Like in the ideal gas mixture without $\chem{N_2}$, we confirm that FHD simulations with the TCR model gives the correct flat structure factor spectra for all values of $P_\chem{N_2}$. 
As shown for the temperature structure factor in Figure~\ref{fig_N2sim}, FHD simulations with the temperature-independent rate constants give incorrect spectra with patterns similar to the simulation without $\chem{N_2}$ (i.e., more significant errors at smaller $\mathbf{k}$).
However, the magnitude of the errors decreases as $P_\chem{N_2}$ increases.
This behavior is consistent with the tendency that the impact of incorrect reaction modelling becomes less significant as the mole fraction of the inert gas increases.

\subsection{\label{subsec_neq_sim_res}Steady State between Isothermal Walls with Different Temperatures}

We now consider an ideal gas mixture of $\chem{NO_2}$ and $\chem{N_2O_4}$ placed between two isothermal walls with different temperatures, $T_1=\SI{320}{\K}$ at $z=0$ and $T_2=\SI{380}{K}$ at $z=L$, and study its fluctuation behavior at steady state.
As described in Section~\ref{subsec_RFHD_implementation}, we use the same values for system size, initial $\chem{NO_2}/\chem{N_2O_4}$ composition, and time step size as the equilibrium case.
To implement isothermal walls~\cite{SrivastavaLadigesNonakaGarciaBell2023}, we use zero concentration flux (Neumann) boundaries for species flux, thermal (Dirichlet) boundaries for heat flux, and slip (Neumann) boundaries for tangential momentum flux along the $z$-direction.
For the $x$- and $y$-directions, we impose periodic boundary conditions.

Before collecting results for time averaging, we wait for the system to reach its steady state by running simulations for $3\times 10^6$ time steps.
This relaxation period was chosen based on the corresponding deterministic simulation; the temperature profile developing along the $z$-direction does not change within machine precision around this time period.
To estimate physical quantities of the steady state as a function of $z$, we compute the cell average of each layer consisting of cells with the same $z$ values at each time step and perform time averaging over $10^7$ time steps.
We denote this average as~$\langle\cdot\rangle_z$.

In this simulation study, we look closely at temperature fluctuations in each layer using the variance $\langle \delta T^2 \rangle_z = \langle [T- \langle T\rangle_z]^2 \rangle_z$, where $T$ is the instantaneous temperature of a cell and $\langle T\rangle_z$ is the mean temperature of the layer.
We compare the values of $\langle \delta T^2 \rangle_z$ with the corresponding local-equilibrium values $\langle \delta T^2 \rangle_{z,eq}$ as given by Eq.~\eqref{eq_Tvar} with $T^{eq} = \langle T\rangle_z$, $\rho^{eq} = \langle \rho\rangle_z$, and $c_{v,mix}$ evaluated at $\langle \rho_1 \rangle_z$ and $\langle \rho_2 \rangle_z$.

To demonstrate the importance of using our thermodynamically-consistent chemistry model in mesoscopic hydrodynamic simulations, we compare the results of our simulation with those of a simulation where reaction rate constants are fixed in each layer throughout the simulation.
More specifically, we compare our TCR model, in which the instantaneous temperature $T$ is used to compute the rate constants $k^\pm(T)$ in each cell (see Eq.~\eqref{eq_kpmT_kpmTst}), to a model in which the rate constants are evaluated at the steady mean temperature $\langle T\rangle_z$ of each layer, i.e., $k^\pm (z) = k^\pm(\langle T\rangle_z)$.
Note that our reaction rate constants $k^\pm(T)$ fluctuate in time around the values close to the mean-profile rate constants $k^\pm(\langle T\rangle_z)$ in the steady state.

\begin{figure}
    \centering
    \includegraphics[width=\linewidth]{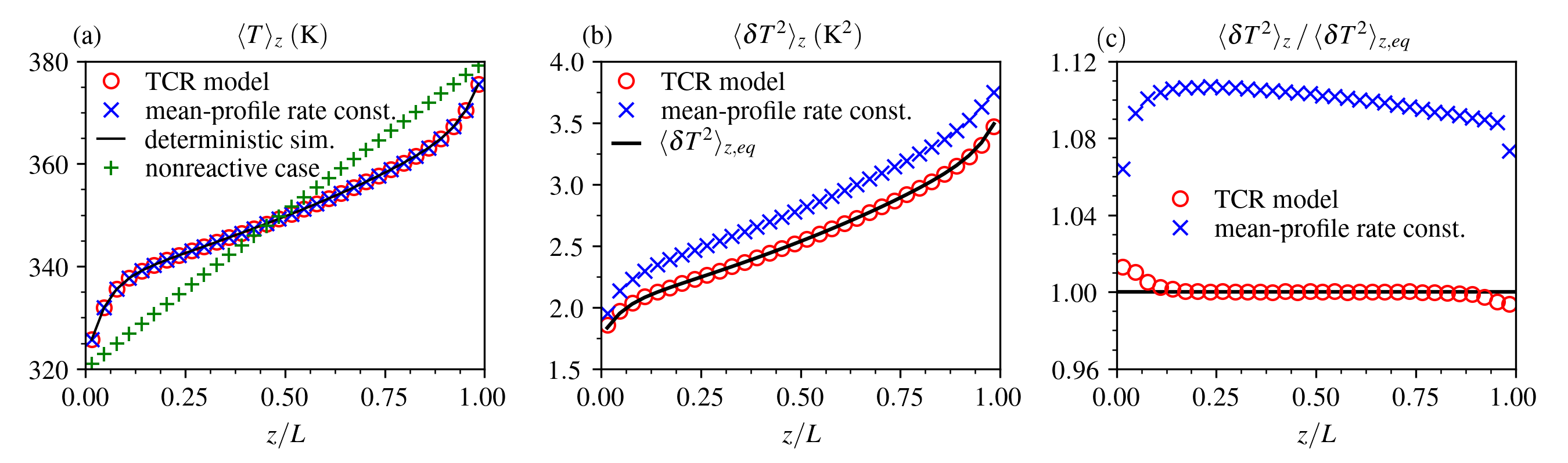}
    \caption{\label{fig_neq_wall_plot1}
    For the steady state of an $\chem{NO_2/N_2O_4}$ mixture placed between two isothermal walls, (a) mean temperature $\langle T\rangle_z$, (b) variance $\langle \delta T^2 \rangle_z$, and (c) the ratio of $\langle \delta T^2 \rangle_z$ to the corresponding local-equilibrium values $\langle \delta T^2 \rangle_{z,eq}$ are plotted as a function of $z$ as red circles.
    Results obtained from the mean-profile rate-constant model are depicted by blue crosses.
    As reference, corresponding results for the deterministic simulation and the local-equilibrium fluctuations are shown in panels~(a) and (b), respectively.
    The temperature profile obtained from the nonreactive FHD simulation is also shown for comparison in panel~(a).}
\end{figure}

In Figure~\ref{fig_neq_wall_plot1}, we compare the profiles of $\langle T\rangle_z$ and $\langle \delta T^2 \rangle_z$ obtained from our TCR model and the mean-profile rate-constant model.
As shown in panel~(a), we first observe that both models produce the same profiles of the mean temperature $\langle T\rangle_z$, which also coincide with that of the corresponding deterministic simulation.
We note that, contrary to the nonreactive case, these temperature profiles $\langle T\rangle_z$ are far from being a linear function in $z$ due to the effect of chemical reactions.
The profiles of $\langle \delta T^2 \rangle_z$ for the two models, as shown in panel~(b), are significantly different.
The mean-profile rate-constant model overall gives larger values of $\langle \delta T^2 \rangle_z$ than our model by about 10 percent.
When compared with the corresponding local-equilibrium fluctuations (i.e., $\langle \delta T^2 \rangle_{z,eq}$), the magnitude of temperature fluctuations observed in our TCR model exhibits minimum deviations.
This can be more clearly seen in panel~(c), where the ratios of $\langle \delta T^2 \rangle_z$ to $\langle \delta T^2 \rangle_{z,eq}$ are plotted.
The values of $\langle \delta T^2 \rangle_z$ observed in our model closely match the corresponding local-equilibrium values in the interior of the domain.  
However, regions near the walls exhibit slight deviations (about up to 2 percent) from the corresponding local-equilibrium case.

We note that in the nonreactive case there is a small contribution to the temperature variance due to the temperature gradient~\cite{GarciaMalekMansourLieCementi1987}. Specifically the change to the variance reaches its peak at $z=L/2$ where
\begin{equation}
    \langle \delta T^2 \rangle_z \approx 
    \langle \delta T^2 \rangle_{z,eq} \left[ 1 + \frac{\Delta x}{L} \left( \frac{T_2 - T_1}{2 T^{eq}}\right)^2 \right] \; .
\end{equation}
For our parameters this correction is $O(10^{-4})$ and, as such, negligible.
In fact, the nonreactive FHD simulation results show that $\langle\delta T^2\rangle_z$ and $\langle\delta T^2\rangle_{z,eq}$ are indistinguishable within statistical errors (see Figure~S3 in the Supplementary Material).

We now investigate the origin of the deviations of the steady-state temperature variance $\langle \delta T^2 \rangle_z$ from the corresponding local-equilibrium temperature variance $\langle \delta T^2 \rangle_{z,eq}$.
We first check whether the deviations of $\langle \delta T^2 \rangle_z$ from $\langle \delta T^2 \rangle_{z,eq}$ can be related to how far chemistry is from equilibrium in each layer.
To this end, we compute the reaction quotient for the dimerization reaction, defined as
\begin{equation}
    Q(\langle[\chem{NO_2}]\rangle_z,\langle[\chem{N_2O_4}]\rangle_z) = 
    \frac{\langle[\chem{N_2O_4}]\rangle_z}{\langle [\chem{NO_2}]\rangle_z^2},
\end{equation}
and compare it with the equilibrium constant $K(\langle T\rangle_z$).
We then interpret these results in terms of entropy production in the system.
For the steady state of a multi-species gas system with zero velocity and body force, the entropy production rate $\mathfrak{v}$ is given by~\cite{Giovangigli2012}
\begin{equation}
\label{eq_entropy_prod_gen}
    \mathfrak{v} = -\frac{\mathbfcal{Q}\cdot\nabla T}{T^2}
    -\sum_s \mathbfcal{F}_s \cdot \nabla\left(\frac{\mu_s}{T}\right)
    -\sum_s \frac{\mu_s M_s \Omega_s}{T}.
\end{equation}
Hence, for our $\chem{NO_2/N_2O_4}$ system, we compute 
\begin{equation}
\label{eq_entropy_prod}
    \mathfrak{v} 
    = -\frac{1}{\langle T\rangle_z^2}\mathbfcal{Q}_z \frac{d}{dz} \langle T \rangle_z
    -\sum_{s=1,2} \mathbfcal{F}_{s,z} \frac{d}{dz} \left(\frac{\mu_s}{\langle T\rangle}_z\right)
    -\frac{(\mu_1-\mu_2) M_1 \Omega_1}{\langle T\rangle_z},
\end{equation}
where $\mathbfcal{Q}_z$, $\mathbfcal{F}_{s,z}$, $\mu_1$, $\mu_2$, and $\Omega_1$ are evaluated at the mean values of the gas state (e.g., $\langle\rho_1\rangle_z$, $\langle\rho_2\rangle_z$, $\langle T\rangle_z$) in each layer.
Note that $\mathbfcal{Q}_z$ has a nonzero constant value and $\mathbfcal{F}_{1,z} = -\mathbfcal{F}_{2,z}$ is a nonzero function of $z$.
This can be seen from the relations, $\nabla\cdot\mathbfcal{F}_{s}=M_s\Omega_s$ and $\nabla\cdot\mathbfcal{Q}=0$, which are obtained from Eqs.~\eqref{eq_FNS_rhos} and \eqref{eq_FNS_rhoE}, respectively, for this nonequilibrium steady state.
In the rest of this section, we investigate the chemical contribution of the entropy production rate $\mathfrak{v}$:
\begin{equation}
     \mathfrak{v}_\mathrm{chem} = -\frac{(\mu_1-\mu_2) M_1 \Omega_1}{\langle T\rangle_z}.
\end{equation}

\begin{figure}
    \centering
    \includegraphics[width=\linewidth]{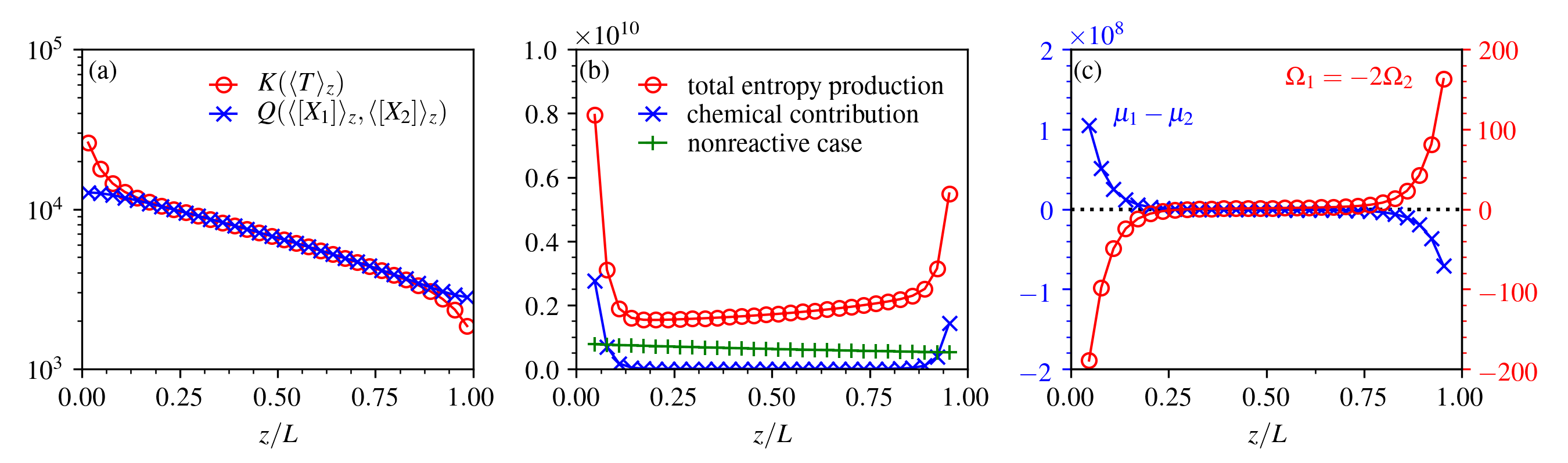}
    \caption{\label{fig_neq_wall_plot2}
    (a) The equilibrium constant $K$ and reaction quotient $Q$ values are compared.
    (b) The total entropy production $\mathfrak{v}$ and its chemical contribution $\mathfrak{v}_\mathrm{chem}$ are shown.
    The entropy production of the nonreactive case is compared.
    (c) The chemical potential difference $\mu_1-\mu_2$ and the molar concentration production rate $\Omega_1$ of $\chem{NO_2}$ are plotted using two different vertical axes.
    All results of the reactive case are obtained from the TCR model.}
\end{figure}

In Figure~\ref{fig_neq_wall_plot2}, we show the aforementioned quantities computed from our TCR model.
In panel~(a), we compare the equilibrium constant values $K(\langle T\rangle_z)$ and the reaction quotient values $Q(\langle[\chem{NO_2}]\rangle_z,\langle[\chem{N_2O_4}]\rangle_z)$.
At the interior of the domain, the $Q$ values are close to the $K$ values.
Near the walls at $z=0$ or $z=L$, the $Q$ values are significantly smaller or greater than the $K$ values, respectively, which implies that chemistry is out of equilibrium.
This observation suggests that the deviations of $\langle \delta T^2 \rangle_z$ from the corresponding local-equilibrium values $\langle \delta T^2 \rangle_{z,eq}$ shown in Figure~\ref{fig_neq_wall_plot1} are related to how far chemistry is from equilibrium in each layer.
The measurement of the entropy production rate shown in panel~(b) confirms this argument.
Since the system is not in thermodynamic equilibrium, the total entropy production rate $\mathfrak{v}$ is positive throughout the system.
However, its chemical contribution $\mathfrak{v}_\mathrm{chem}$ is zero in the interior and positive near the walls.
The profiles of $\mu_1-\mu_2$ and $\Omega_1=-2\Omega_2$ shown in panel~(c) are consistent with the profiles of the $K$-$Q$ pair and $\mathfrak{v}_\mathrm{chem}$. 
While both $\Omega_1<0$ and $K>Q$ imply that the dimerization reaction is dominant near the wall at $z=0$, both $\Omega_1>0$ and $K<Q$ imply that the dissociation reaction is dominant near the wall at $z=L$.
In both cases, the chemical entropy production rate $\mathfrak{v}_\mathrm{chem}$ is positive because $\mu_1-\mu_2$ and $\Omega_1$ are of opposite signs.
We note that the entropy production profile in the absence of the dimerization reaction is significantly different.
In the nonreactive case shown in Figure~\ref{fig_neq_wall_plot2}(b), the sharp increase in the total entropy production rate $\mathfrak{v}$ near the walls is not observed and the magnitude of $\mathfrak{v}$ is notably smaller even in the interior of the system.
For the decomposition of $\mathfrak{v}$ (see Eq.~\eqref{eq_entropy_prod}) in both reactive and nonreactive cases, see Figure~S4 in the Supplementary Material.

\section{\label{sec_conclusion}Conclusion}

In this paper, we have formulated a stochastic chemistry model, called the thermodynamically-consistent reaction (TCR) model, that can be incorporated into a mesoscopic hydrodynamic description to construct a thermodynamically-consistent stochastic simulation method for reactive gas mixtures. 
Thermodynamic consistency requires that all elementary reactions be reversible and that the forward and reverse reactions be consistent with thermodynamic equilibrium.  
To guarantee this property, for each reversible reaction pair we specify only one of the reactions and determine the other from Eq.~\eqref{eq_Keq_kpm}, consistent with standard practice (see, for example, Refs.~\citenum{Giovangigli2012, Keizer1987, KeeRupleyMeeksMiller1996}).
An essential requirement of the TCR model is that reaction rates must be evaluated with the instantaneous temperature.
Even if temperature fluctuations are rather small, replacing these fluctuating rate constants by temperature-independent (for equilibrium) or mean-profile (for nonequilibrium steady state) rate constants results in an incorrect fluctuation spectrum.

By incorporating the TCR model into the fluctuating hydrodynamics (FHD) description for ideal gas mixtures, we have constructed a reactive FHD simulation method for reactive gas mixtures.
By performing equilibrium and nonequilibrium simulations of $\chem{NO_2}/\chem{N_2O_4}$ gas mixtures undergoing a reversible dimerization reaction, we confirmed the aforementioned features of the TCR model as essential requirements for the fluctuations of the system to be captured correctly.
For our simulation study, we considered the case where each chemical species has constant heat capacity in the temperature range of interest and formulated the constant-heat-capacity TCR model.
As described below, this simpler TCR model has interesting features that can reduce the complexity of computational modeling of reactive gas mixture.
It also enabled us to analyze the fluctuations behavior of the equilibrium system via the static structure factor spectra obtained analytically using the linearized FHD model given as linear stochastic PDEs (see Appendix~\ref{appendix_structfac}).

In this paper, we have focused on the constant-heat-capacity TCR model with forward rate given in a modified Arrhenius form. 
For this case, the temperature dependence of the equilibrium constant $K_r(T)$ is then completely characterized by three parameters, $K_r(T^{st})$, $A_r$, and $B_r$, as appear in Eq.~\eqref{eq_KT_KTst}.
Since this form is identical to the modified Arrhenius equation~\eqref{eq_kpmT_kpmTst}, if one of the rate constants is modeled by the latter equation, so is the other rate constant.
This feature simplifies thermodynamically-consistent reaction modeling into the task of determining one set of parameters, $k_r^\pm(T^{st})$, $\alpha_r^\pm$, and $\beta_r^\pm$, from chemical kinetics databases and computing the other set using Eqs.~\eqref{eq_Keq_kpm} and \eqref{eq_CHC_TCR_param_rel}.
For the example of the reversible dimerization of $\chem{NO_2}$, the overall modeling procedure was described in Section~\ref{subsec_reaction_params}.
Note that the constant-heat-capacity assumption is at least qualitatively valid unless there are dramatic temperature changes in a reactive gas system.

The TCR model can easily be generalized to reactive gas simulations using real thermochemistry data with detailed temperature dependencies.
For more realistic systems, the heat capacity of each species is specified  as a function of temperature.  High-fidelity fits of this data are available from a number of sources such as the NIST Chemistry WebBook~\cite{NISTchemistrywebbook, LinstromMallard2001}.
Thermodynamic quantities such as entropy, enthalpy, and chemical potential for each species can then be readily computed.
For each reversible reaction pair $r$ one of the rate constants, either $k_r^+(T)$ or $k_r^-(T)$ (typically the forward rate) needs to be given as a function of temperature $T$.
The other rate constant must then be determined via Eq.~\eqref{eq_Keq_kpm} using the equilibrium constant $K_r(T)$, which can be computed from the chemical potentials.
Rate constants are typically given in the form of the modified Arrhenius equation, see, for example, the NIST chemical kinetics database~\cite{NISTchemkin} or CHEMKIN~\cite{KeeRupleyMeeksMiller1996}. 
In general, the reverse rate constant computed using
Eq.~\eqref{eq_Keq_chempot} need not be in modified Arrhenius form.
Note that the mesoscopic hydrodynamic model employed must be based on the same thermodynamic model (or, the same heat capacity function $c_{p,s}(T)$ for each species $s$) in order to achieve thermodynamic consistency.

In Section~\ref{subsec_eq_sim_res}, we demonstrated that in general the use of temperature-independent rate constants in hydrodynamic simulations leads to thermodynamic inconsistency.
However, when the specific internal energies of the chemical species are identical, the system is thermodynamically consistent, as shown in Appendix~\ref{appendix_structfac}.
In this ``color'' chemistry model where chemical species are thermally indistinguishable, the heat of reaction is zero~\cite{NicolisPrigogine1977, BarasMalekMansour1997}.

The measurement of fluctuations is increasingly being used to study molecular relaxation and chemical reaction mechanisms.
These measurements are done in both laboratory experiments (e.g., light scattering) and molecular simulations~\cite{BlumSalsburg1968, BrunoGiovangigli2022}.
In some cases the fluctuation spectra are measured in nonequilibrium systems, such as flame fronts and shocks~\cite{RoyPicardoEmersonLieuwenSujith2023}.
The quantitative analysis of such data requires accurate formulations of the underlying physical and chemical processes.
The combination of fluctuating hydrodynamics with a thermodynamically-consistent reaction model will be a valuable diagnostic tool in this context.

\section*{Supplementary Material}
Numerical implementation of reactive FHD; Simulation results for equilibrium system without reactions, equilibrium system with temperature-independent rate constants, nonreactive nonequilibrium system with isothermal walls, and the decomposition of 
 entropy production rate for reactive nonequilibrium system with isothermal walls.

\begin{acknowledgments}
This work was supported in part by the National Science Foundation under Grant No.\ CHE-2213368 and Grant No.\ DMS-1840265. 
This work was also supported in part by the U.S.\ Department of Energy, Office of Science, Office of Advanced Scientific Computing Research, Applied Mathematics Program under Contract No.\ DE-AC02-05CH11231. 
\end{acknowledgments}

\appendix

\section{\label{appendix_formul_p}Chemistry Formulation Based on Partial Pressures}

In Section~\ref{sec_TCR}, our thermodynamically-consistent chemistry formulation is presented in terms of molar concentrations $[X_s]$.
Equivalently, it can be given in terms of partial pressures $P_s = [X_s] RT$.
In this section, we summarize the corresponding expressions for our main results. 
Here we assume that the equilibrium constant $\overline{K}_r$ and rate constants $\overline{k}_r^\pm$ are defined in terms of $P_s/P^{st}$ so that
\begin{align}
    &\overline{K}_r = \prod_{s=1}^{\Ns} \frac{(P_s/P^{st})^{\nu_{sr}^-}}{(P_s/P^{st})^{\nu_{sr}^+}}\quad
    \mbox{at equilibrium}, \\
    &a_r^\pm = \overline{k}_r^\pm \prod_{s=1}^{\Ns} (P_s/P^{st})^{\nu_{sr}^\pm}.
\end{align}
Note that these partial pressure based quantities are related to the corresponding molar concentration based quantities as
\begin{equation}
    \overline{K}_r(T) = K_r(T)\left(\frac{RT}{P^{st}}\right)^{\Delta n_r},\quad
    \overline{k}_r^\pm(T) = k_r^\pm(T)\left(\frac{RT}{P^{st}}\right)^{-n_r^\pm},
\end{equation}
where $\Delta n_r = \sum_{s=1}^{\Ns} \nu_{sr}$ and $n_r^\pm = \sum_{s=1}^{\Ns} \nu_{sr}^\pm$.

When partial pressures are employed instead of molar concentrations, it is more convenient to use $\overline{\mu}_s^\circ (T)$ instead of $\mu_s^\circ(T)$, see Eq.~\eqref{eq_mus_XT}.
Note that these quantities satisfy
\begin{equation}
\label{musPmus}
    \overline{\mu}_s^\circ (T) = \mu_s^\circ(T) + \frac{RT}{M_s}\log \frac{P^{st}}{RT}.
\end{equation}
In the resulting expressions for $\overline{K}_r$ and $\overline{k}_r^\pm$ for the general case considered in Section~\ref{subsec_TCR_principles}, the per-particle chemical potential $\hat{\mu}_s^\circ (T)$ is replaced by $\hat{\overline{\mu}}_s^\circ (T) = (M_s/RT) \overline{\mu}_s^\circ (T)$. 
Specifically, Eqs.~\eqref{eq_Keq_chempot} and \eqref{eq_kpm_lambda} become respectively
\begin{align}
\label{KrP}
    &\overline{K}_r(T) = \exp\left(-\sum_{s=1}^{\Ns} \nu_{sr} \hat{\overline{\mu}}_s^\circ (T) \right),\\
\label{}
    &\overline{k}^\pm_r(T) = \lambda_r(T) \exp\left(\sum_{s=1}^{\Ns}\nu_{sr}^\pm \hat{\overline{\mu}}_s^\circ(T)\right).
\end{align}

For the constant-heat-capacity TCR model considered in Section~\ref{subsec_const_heat_cap_TCR_model}, Eq.~\eqref{eq_muhatT_muhatTst} becomes
\begin{equation}
    \hat{\overline{\mu}}_s^\circ(T) - \hat{\overline{\mu}}_s^\circ(T^{st}) 
    = \frac{M_s \epsilon_s}{R}\left(\frac{1}{T}-\frac{1}{T^{st}}\right)
    - \hat{c}_{p,s} \log \frac{T}{T^{st}},
\end{equation}
where $\hat{c}_{p,s} = (M_s/R) c_{p,s}$.
Thus, Eq.~\eqref{eq_KT_KTst} becomes
\begin{equation}
    \overline{K}_r (T) = \overline{K}_r (T^{st}) \exp\left[-\frac{\overline{A}_r}{R}\left(\frac{1}{T}-\frac{1}{T^{st}}\right)\right]
    \left(\frac{T}{T^{st}}\right)^{\overline{B}_r},
\end{equation}
where $\overline{A}_r = A_r$ and $\overline{B}_r = B_r + \Delta n_r$.
Finally, when the rate constants are given in the form of the modified Arrhenius equation, 
\begin{equation}
    \overline{k}_r^\pm (T) = \overline{k}_r^\pm (T^{st}) \exp\left[-\frac{\overline{\alpha}_r^\pm}{R}\left(\frac{1}{T}-\frac{1}{T^{st}}\right)\right]
    \left(\frac{T}{T^{st}}\right)^{\overline{\beta}_r^\pm},
\end{equation}
the parameters $\overline{\alpha}^\pm$ and $\overline{\beta}^\pm$ must satisfy 
\begin{equation}
    \overline{\alpha}_r^- = \overline{\alpha}_r^+ - \overline{A}_r, \quad
    \overline{\beta}_r^- = \overline{\beta}_r^+ - \overline{B}_r,
\end{equation}
for thermodynamic consistency.

\section{\label{appendix_structfac}Structure Factor Analysis for Dimerization}

For a one-dimensional two-species ideal gas mixture undergoing a reversible dimerization reaction at equilibrium, we show that the stochastic chemistry formulation based on our TCR model gives the correct flat structure factor spectra as predicted by the equilibrium statistical mechanics.
Following the fact that adding chemical reactions should not change the thermodynamic equilibrium state of the system, we show that the correct equilibrium structure factor spectra established by nonreactive FHD are not disturbed by the inclusion of our chemistry model.
Here, we denote transpose and complex transpose by $(\cdot)^T$ and $(\cdot)^\ast$, respectively.

We first consider the nonreactive system.
For our purposes, let $\mathbf{U} = \left[ \delta\rho_1, \delta\rho_2, v, \delta T \right]^T$, where $\delta \rho_s = \rho_s - \rho_s^{eq}$ and $\delta T = T - T^{eq}$.
Like a one-species system~\cite{DonevVandenEijndenGarciaBell2010}, the linearized FHD equations for a two-species system can be written as
\begin{equation}
\label{nonreactlinFHD}
    \frac{\partial}{\partial t}\mathbf{U} = 
    -\frac{\partial}{\partial x}\left(\mathbf{A}_1\mathbf{U}\right)
    +\frac{\partial^2}{\partial x^2}\left(\mathbf{A}_2\mathbf{U}\right)
    +\frac{\partial}{\partial x}\left(\mathbf{B}\mathbfcal{Z}\right),
\end{equation}
where the first, second, and third terms in the right-hand side correspond to hyperbolic, dissipative, and stochastic fluxes, respectively, and $\mathbfcal{Z}$ is a collection of independent standard Gaussian white noise fields.
Since transport coefficients are dependent on composition~\cite{BellGarciaWilliams2010}, it is possible but tedious to derive explicit expressions of $\mathbf{A}_1$, $\mathbf{A}_2$, and $\mathbf{B}$. 
Note, however, that we do not need explicit expressions.
Instead, we assume that the linearized FHD equations~\eqref{nonreactlinFHD} guarantees thermodynamic consistency.
In other words, the following equation~\cite{DonevVandenEijndenGarciaBell2010}, from which the structure factors $\mathbf{S}(k) = V\langle \widehat{\mathbf{U}}_k \widehat{\mathbf{U}}_k^\ast\rangle$ can be determined,
\begin{equation}
\label{Seqnnonreact}
    \left(-ik\mathbf{A}_1-k^2\mathbf{A}_2\right)\mathbf{S}(k)
    + \mathbf{S}(k)\left(ik\mathbf{A}_1^T-k^2\mathbf{A}_2^T\right)
    + \mathbf{B}\mathbf{B}^T = 0,
\end{equation}
is satisfied by the correct structure factors $\mathbf{S}^{eq}$, which are obtained from the equilibrium statistical mechanics~\cite{BellGarciaWilliams2010, BalakrishnanGarciaDonevBell2014}:
\begin{equation}
\label{Scorrect}
    \mathbf{S}^{eq} = \mathrm{diag}\left(
    \frac{M_1}{N_A}\rho_1^{eq}, \frac{M_2}{N_A}\rho_2^{eq},\frac{\kb T^{eq}}{\rho_1^{eq}+\rho_2^{eq}},
    \frac{\kb (T^{eq})^2}{c_{v,1}\rho_1^{eq}+c_{v,2}\rho_2^{eq}}
    \right).
\end{equation}

We now consider the reactive system undergoing dimerization.
By using $M_1\Omega_1 = -M_2\Omega_2$ and expressing $\partial (\delta T)/\partial t$ in terms of $\partial \rho_s /\partial t$ and $\partial (\rho E)/\partial t$, we write the linearized FHD equations as
\begin{equation}
\label{dUdtreact}
    \frac{\partial}{\partial t}\mathbf{U} = 
    -\frac{\partial}{\partial x}\left(\mathbf{A}_1\mathbf{U}\right)
    +\frac{\partial^2}{\partial x^2}\left(\mathbf{A}_2\mathbf{U}\right)
    + \mathbf{A}_0\mathbf{U}
    +\frac{\partial}{\partial x}\left(\mathbf{B}\mathbfcal{Z}\right)
    +\mathbf{b}z.
\end{equation}
Here, $z$ is a standard Gaussian white noise field and the additional terms are given as
\begin{equation}
    \mathbf{A}_0\mathbf{U} = M_2 \overline{\Omega}_2
    \begin{bmatrix}
        -1 \\ 1 \\ 0 \\ \phi
    \end{bmatrix},
    \quad
    \mathbf{b}z =  M_2 \widetilde{\Omega}_2
    \begin{bmatrix}
        -1 \\ 1 \\ 0 \\ \phi
    \end{bmatrix},
\end{equation}
where
\begin{equation}
    \phi = \frac{(\epsilon_1-\epsilon_2) + (c_{v,1}-c_{v,2})T^{eq}}{c_{v,1}\rho_1^{eq}+c_{v,2}\rho_2^{eq}}.
\end{equation}
To derive explicit expressions of $\mathbf{A}_0$ and $\mathbf{b}$, we linearize the rate constants,
\begin{equation}
    k^\pm(T) = k^\pm_{eq}\left\{1+\left(\frac{\alpha^\pm}{RT^{eq}}+\beta^\pm\right)\frac{\delta T}{T^{eq}}\right\},
\end{equation}
where $k^\pm_{eq}=k^\pm(T^{eq})$, and obtain
\begin{align}
    M_2\overline{\Omega}_2 
    &= k^-_{eq}\rho_2^{eq}\left\{
        2\frac{\delta\rho_1}{\rho_1^{eq}}
        -\frac{\delta\rho_2}{\rho_2^{eq}}
        +\psi\frac{\delta T}{T^{eq}}
    \right\},\mbox{  where } \psi = \frac{A}{RT^{eq}}+B, \\
    M_2\widetilde{\Omega}_2 
    &= \sqrt{2\frac{M_2}{N_A}k_{eq}^-\rho_2^{eq}}\; z.
\end{align}
Thus, we finally obtain
\begin{equation}
\label{A0b}
    \mathbf{A}_0 = k_{eq}^-\rho_2^{eq}
    \begin{bmatrix}
        -\frac{2}{\rho_1^{eq}} & \frac{1}{\rho_2^{eq}}  & 0 & -\frac{\psi}{T^{eq}} \\
        \frac{2}{\rho_1^{eq}}  & -\frac{1}{\rho_2^{eq}} & 0 & \frac{\psi}{T^{eq}}  \\
        0 & 0 & 0 & 0 \\
        \frac{2\phi}{\rho_1^{eq}}  & -\frac{\phi}{\rho_2^{eq}} & 0 & \frac{\phi\psi}{T^{eq}}
    \end{bmatrix},
    \quad
    \mathbf{b} = \sqrt{2\frac{M_2}{N_A}k_{eq}^-\rho_2^{eq}}
    \begin{bmatrix}
        -1 \\ 1 \\ 0 \\ \phi
    \end{bmatrix}.
\end{equation}

The structure factors $\mathbf{S}(k)$ resulting from Eq.~\eqref{dUdtreact} can be determined by
\begin{equation}
\label{Seqnreact}
    \left(\mathbf{A}_0-ik\mathbf{A}_1-k^2\mathbf{A}_2\right)\mathbf{S}(k)
    + \mathbf{S}(k)\left(\mathbf{A}_0^T+ik\mathbf{A}_1^T-k^2\mathbf{A}_2^T\right)
    + \mathbf{B}\mathbf{B}^T + \mathbf{b}\mathbf{b}^T= 0.
\end{equation}
Hence, if the following condition is satisfied by $\mathbf{S}^{eq}$, $\mathbf{A}_0$, and $\mathbf{b}$,
\begin{equation}
\label{condA0b}
    \mathbf{A}_0 \mathbf{S}^{eq} + \mathbf{S}^{eq} \mathbf{A}_0^T + \mathbf{b}\mathbf{b}^T= 0,
\end{equation}
by combining that Eq.~\eqref{Seqnnonreact} is satisfied by $\mathbf{S}^{eq}$, one can show that $\mathbf{S}^{eq}$ satisfies Eq.~\eqref{Seqnreact}.
By using explicit expressions of $\mathbf{S}^{eq}$, $\mathbf{A}_0$, and $\mathbf{b}$ (see Eqs.~\eqref{Scorrect} and \eqref{A0b}), one can easily show that Eq.~\eqref{condA0b} holds.
Therefore, our TCR chemistry formulation gives the correct structure factors.

We have a couple of remarks for this analysis.
First, by comparing Eq.~\eqref{Seqnnonreact} (nonreactive case) and Eq.~\eqref{Seqnreact} (reactive case), one can predict that errors in the structure factors appear at small $k$ values if a thermodynamically-inconsistent chemistry formulation is used.
This is because the contribution of the additional reactive terms containing $\mathbf{A}_0$ and $\mathbf{b}$ becomes less significant in Eq.~\eqref{Seqnreact} for larger $k$ values and the two equations becomes the same asymptotically in the large $k$ limit.  
Second, using the definitions of $A$ and $B$ (see Eq.~\eqref{eq_Ar_Br_def}), one can show that
\begin{equation}
    \psi = -\frac{M_2}{RT^{eq}}\left(c_{v,1}\rho_1^{eq}+c_{v,2}\rho_2^{eq}\right)\phi.
\end{equation}
Hence, if $\epsilon_1+c_{v,1}T^{eq} = \epsilon_2+c_{v,2}T^{eq}$, that is, the specific internal energies of $\chem{A}$ and $\chem{A_2}$ coincide, both $\psi$ and $\phi$ become zero.
In this special case, the temperature dependence in $\mathbf{A}_0$ and $\mathbf{b}$ disappears but thermodynamic consistency is still maintained.
Similarly, simple models, such as the Brusselator, that use ``color chemistry'' (species are thermally indistinguishable) are themodynamically consistent despite using constant rates~\cite{NicolisPrigogine1977, BarasMalekMansour1997}.

%


\end{document}